\documentstyle[epsfig]{mn}
 at 10truept
\title
     [ ]
{\vglue-3.0truecm
\centerline{\it\small For submission to Monthly Notices}
\vglue 2.5truecm
    Prospects for galaxy-mass relations from the 6dF Galaxy Survey.
\author
     [ D. Burkey \& A.N. Taylor]
     { D. Burkey \& A.N. Taylor\\
     Institute for Astronomy,
     University of Edinburgh,
     Royal Observatory,
     Blackford Hill,
     Edinburgh,
     U.K.\\
        email:db@roe.ac.uk, ant@roe.ac.uk}}
\def\bib{\parskip=0pt\par\noindent\hangindent\parindent
    \parskip =2ex plus .5ex minus .1ex}

\newcommand{\be}{\begin{equation}}
\newcommand{\ee}{\end{equation}}
\newcommand{\ba}{\begin{eqnarray}}
\newcommand{\ea}{\end{eqnarray}}

\newcommand{\nn}{\nonumber \\}
\newcommand{\nnb}{\begin{displaymath}}
\newcommand{\nne}{\end{displaymath}}
\newcommand{\de}{\partial}

\newcommand{\r}{\mbox{\boldmath $r$}}
\newcommand{\rhatb}{\mbox{\boldmath $\hat{r}$}}

\newcommand{\C}{\mbox{\boldmath $C$}}
\newcommand{\k}{\mbox{\boldmath $k$}}
\newcommand{\vb}{\mbox{\boldmath $v$}}

\newcommand{\hMpc}{\,h^{-1}{\rm Mpc}}

\newcommand{\khat}{\hat{\k}}
\newcommand{\Mpc}{\, h^{-1}{\rm Mpc}}
\newcommand{\Mpch}{\, h{\rm Mpc}^{-1}}

\newcommand{\eF}{{\cal F}}
\newcommand{\sky}{{\rm sky}}
\newcommand{\eff}{{\rm eff}}

\newcommand{\thetab}{\mbox{\boldmath $\theta$}}
\newcommand{\rgl}{\rangle}
\newcommand{\lgl}{\langle}

\newcommand{\Cov}{{\rm Cov}}

\def\bib{\parskip=0pt\par\noindent\hangindent\parindent
    \parskip =2ex plus .5ex minus .1ex}

\begin{document}
\bibliographystyle{astron}
\maketitle


\begin{abstract}
We develop new methods to study the properties
of galaxy redshift surveys and radial peculiar velocity surveys,
both individually and combined. We derive the Fisher information
matrix for redshift surveys, including redshift distortions and
stochastic bias. We find exact results for estimating the
marginalised accuracy of a two-parameter measurement of the
amplitude of galaxy clustering, $A_g$, and the distortion
parameter $\beta$. The Fisher matrix is also derived for a radial
peculiar velocity survey and we discuss optimisation of these
surveys for equal timescales. The Fisher Super-Matrix, combining
both surveys, is derived. We apply these results to investigate
the 6 degree Field Galaxy Survey (6dFGS), currently
underway on the UK Schmidt Telescope (UKST). The survey will
consist of $\sim 10^{5}$ K-band selected galaxies with redshifts
and a subset of $\sim 15000$ galaxies with radial peculiar
velocities. We find for the redshift survey that we can measure the three
parameters $A_g$, $\Gamma$ and $\beta$ to about $3\%$ accuracy,
but will not be able to detect the baryon abundance, or the
matter-galaxy correlation coefficient, $r_g$. The peculiar velocity
survey will jointly measure the velocity amplitude $A_v$ and
$\Gamma$ to around $25\%$ accuracy. A conditional estimate of
the amplitude 
$A_v$ alone, can be made to $5\%$. When the surveys are combined however, the major
degeneracy between $\beta$ and $r_g$ is lifted and 
we are able to measure $A_g$, $\Gamma$, 
$\beta$ and $r_g$ all to the $2\%$ level, significantly improving on
current estimates. Finally we consider scale dependence of $r_g$ and
the biassing parameter $b$. We find that measurements for these
averaged over logarithmic passbands can be constrained to the level of
a few percent. 
\end{abstract}


\section{introduction}

The quest to constrain cosmological parameters has been boosted in recent
years by the emergence of large data sets from CMB experiments and galaxy
surveys. In particular the  first data release from WMAP combined with
information from the 2 degree field (2dF) redshift survey and Lyman
$\alpha$ forest data has allowed many cosmological parameters to be
tightly constrained (Spergel et al. 2003). The Sloan Digital Sky
Survey (SDSS) when complete, will constitute a fourfold increase in
redshifts over 2dF. The 6 degree Field (6dF) galaxy survey (Wakamatsu, 2003,
Colless,
1999) offers a
unique combination of a wide field redshift survey with a homogeneous
subset of peculiar velocities. It is the aim of this paper, in the
context of the present data-boom, to predict the unique advantages of
the 6dF galaxy survey over its contemporaries.

Galaxy surveys have long been an invaluable source of information
in cosmology, allowing cosmologists to infer the large-scale
clustering of matter in the Universe, and cosmological parameters.
But despite their prominent role in our understanding of the
clustering of matter, galaxy redshift surveys are fundamentally
limited as a probe as we do not have a complete theory of galaxy
formation from first principles. This limitation is characterised
by the galaxy bias parameter, which is an unknown function
relating the galaxy distribution to that of matter. In practice
this function will be determined observationally, but its
structure will help shape the theory of galaxy formation.

Galaxy redshift surveys also suffer from the well-known distortion
due to the use of  Doppler redshifts to infer distances. But
redshifts are degenerate with peculiar velocities, the deviations
from the Hubble flow generated by gravitational instability~(Peebles, 1980). This projection of radial velocity information
into redshift surveys provides aditional complication to the
analysis of the clustering pattern, but also injects extra
information on the dynamics of large-scale structure which couples
directly to the mass distribution.

In the local universe, the best way to probe the matter
distribution is from the local peculiar velocity field. At higher
redshift gravitational lensing becomes the method of choice for
probing the distribution of matter, while on even higher redshifts
the CMB provides the only viable and accurate method.

By combining redshifts with distance indicators, such as the
$D_{n}-\sigma$ relation, cosmologists have been able to estimate
both true distance and radial peculiar velocities for local
galaxies. Since estimating the true distance is a complicated, and
time consuming process, these surveys have tended to be much
smaller in size than the redshift surveys. In addition the
distance estimators are subject to much larger uncertainties and
biases than are the redshift estimates, and so have suffered from
large uncertainties.

 One of the problems with this programme has been the difficulty
 in collecting a large homogeneous sample. Progress in this field
 has been limited by the need to patch together a number of
 surveys with almost inevitable systematic differences. The subject of
 the different biasses which plague peculiar velocity surveys is
 reviewed in Strauss and Willick (1995). There are two fundamental
 problems. $D_n$-$\sigma$ and Tully Fisher relations derived from
 different data will be systematically different, and a relation
 derived from one sample cannot be applied to another. Secondly, the
 observed quantities such as the velocity dispersions and apparent
 magnitudes needed for $D_n$-$\sigma$, will differ between surveys
 since each survey uses different observational methods and applies
 different corrections. The Mark III catalogue is one such compilation
 for which analysis has been done including the POTENT reconstruction
 where the 3-dimensional velocity is reconstructed from a potential
 - e.g. Kolatt et al. (2000). Another catalogue comprised of $\sim 1600 $
 field galaxies is the SFI catalogue. Both SFI and Mark III have been
 used to analyse the density power spectrum to constrain cosmological
 parameters as for example in Zehavi and Dekel (2000) .

In the age of mega-surveys this problem of compiling inhomogeneous
 surveys  should be
 alleviated by large coherent galaxy samples with accurate
 distance indicators.
The 6dF Galaxy Survey
is the first such data set. The 6dFGS is currently underway on
the UK Schmidt Telescope (UKST). The survey will consist of $\sim
10^{5}$ K-band selected galaxies with redshifts and a subset of
$\sim 15000$ of the brightest galaxies with radial peculiar
velocities. The galaxies are sampled from the K-band 2 Micron All Sky
Survey (2MASS) 
Extended Source Catalogue~(Skrutskie, 2000) and so are dominated by early type galaxies. The advantage
of the K-band is that it selects light from the old stellar
population, and so presumably is a good indicator of the mass of
the galaxy as a whole.

In addition to redshifts the 6dF will also collect galaxy
distances. The 6dFGS will be the first combined
redshift and velocity survey, with the advantage that the
selection criteria for both surveys are matched. The prospect of a
homogeneous and integrated galaxy redshift and radial velocity
survey is a big step forward in the analysis of peculiar
velocities, which have suffered from both sampling and
inhomogeneity effects.

In this paper we develop an information-theory analysis of
galaxy redshift surveys and radial peculiar velocity surveys, both
individually and combined. Such an analysis is required both for
survey design and to understand the sensitivity of the survey to
cosmological parameters and so help determine which parameters the
survey can be optimised to measure. These methods are general to
the construction and analysis of galaxy redshift and radial
velocity surveys, and can readily be applied to, e.g., the Sloan
Digital Sky Survey (SDSS) dataset or the radial velocity fields
probed by galaxy clusters from Sunyaev-Zel'dovich surveys of
the CMB, such as in the case of the Planck survey (Lawrence and Lange 1997).

The paper is laid out as follows. In Section 2 we describe our
information theory methods, based on the Fisher information
matrix. In Section 3 we derive the results we shall need for the
analysis of galaxy redshift surveys while in Section 4 we apply
these results to the problem of optimising their design. In Section
5 we estimate the sensitivity of a galaxy redshift survey to
cosmological parameters, using the 6dF as our fiducial model. The
analytic results we require for an analysis of the radial
peculiar velocity field are derived in Section 6, and used to
optimise such surveys in Section 7. Parameter estimation from a
peculiar velocity survey is studied in Section 8, while in Section 9 we analyse the
combined surveys. We present our conclusions in Section 10. There
are also two appendices, with Appendix A deriving a series of
useful formulae for the analysis of redshift surveys, and Appendix
B calculating the bivariate Fisher matrix. We begin with a review of
the Fisher information matrix.

\section{Fisher Information Analysis}

\subsection{The single field case}

The Fisher information matrix measures the information content of
a random field, $\phi(\r)$, about a set of parameters, $\thetab$,
that characterise that field. If this field is Gaussian
distributed, the Fisher matrix is~(Vogeley and Szalay 1996; Tegmark,
Taylor and Heavens 1997, Tegmark 1997; Taylor and Watts 2001), 
 \be
    \eF_{ij}= \frac{1}{2} \int\! \frac{d^3 \!k}{(2 \pi)^3}\,
    \de_i \ln P_{\phi \phi}(\k) \,
    \de_j \ln P_{\phi \phi}(\k) \, V_{\rm eff}(\k)
 \label{fish1}
 \ee
 where
 \be
    \lgl \phi(\k) \phi^*\!(\k) \rgl = (2 \pi)^3 P_{\phi \phi}(k)
    \delta_D(\k-\k')
 \ee
 defines the power spectrum of a homogeneous field $\phi$, and we have assumed
 $\lgl \phi \rgl=0$. The effective volume of the survey is
 \be
        V_{\rm eff}(\k) =
        \int_V \! d^3\!r \,\left(\frac{P_{\phi\phi}(\k)}{
        P_{\phi\phi}(\k)+N(\r)}\right)^2
 \ee
 where the spatial integral is taken over the volume sampling the
 $\phi$-field.
 As the field is
 Gaussian all the information characterising the field is
 contained in the power spectrum,
 $P_{\phi \phi}(\k)=P_{\phi \phi}(\k|\thetab)$. The gradients
 in equation (\ref{fish1}), $\de_i \equiv \de/\de\theta_i$, are
 taken in parameter space.

 If we randomly sample the field by a set of discrete points,
 the covariance of the sampled field is
 \be
    C_{\phi \phi}(\k,\r) = 
    [P_{\phi \phi}(\k) + N(\r)]
 \ee
 where $N(\r)$ is a noise term that may vary spatially.

 The Fisher information matrix contains both the conditional
 error on a parameter on its diagonal and the marginalised error,
 \be
   \lgl  \Delta \theta^2_{i} \rgl_{\rm marg} =
   [\eF^{-1}]_{ii} \ge 1/\eF_{ii},
 \ee
 for the $i^{th}$ parameter.
The Cram{\'e}r-Rao inequality~(Kendall and Stuart 1969),
 \be
    \lgl \Delta \theta_i \Delta \theta_j \rgl \ge [\eF^{-1}]_{ij}
 \ee
 where $\Delta \thetab = \thetab-\thetab_0$ is the deviation of
 the parameters from their true value $\thetab_0$,
 guarantees that the maginalised error is the smallest possible
 uncertainty on a parameter from a given measurement, and hence
 gives the minimum variance bound on marginalised parameter
 estimates. Furthermore the Cram{\'e}r-Rao bound is an equality if
 maximum likelihood methods are applied.


If the parameter is the power itself the uncertainty on a
measurement of the power, band averaged over a logarithmic
passband of width $d \ln k$, is given by~(Feldman, Kaiser and Peacock 1994)
 \be
    \Delta P(k) = \frac{2 \pi P(k)}
    {\sqrt{k^3 d \ln k V_{\rm eff}(k)}}.
    \label{bandpow}
 \ee

\subsection{Multiple Fields}

Multiple fields can be incorporated into the Fisher Matrix
formalism by generalising the definition of the data vector and
its covariance matrix. Assuming we have two fields, $\phi_1$ and
$\phi_2$, we can form a covariance super-matrix
 \be
    \C  =
     \left( \begin{array}{cc} C_{11} & C_{12} \\
                                 C_{21} & C_{22} \end{array}
                                                        \right)
 \ee
 where each of the individual covariances
 $ C_{\alpha \beta} = \lgl \phi_\alpha \phi_\beta \rgl$ where
 $\alpha$, $\beta=1,2$
 labels the fields. This super-matrix can be used to generate a Fisher
 super-matrix (eg Zaldarriaga, Spergel and Seljak 1997, Bouchet,
Prunet and Sethi 1999),
\begin{equation}
{\cal F}_{ij} = \int d^3 k d^3 r \sum_{{\rm XY}} \partial_{i}
C_{{\rm X}} [{\rm Cov} ( C_{{\rm X}} C_{{\rm Y}}) ]^{-1} \partial_{j}
C_{{\rm Y}}.
\label{super}
\end{equation}
 The full expression of the
 Fisher super-matrix is given in the Appendix by equation (\ref{supermat}).

\subsection{Parameter correlation coefficient}

As well as the uncertainty on a measurement of a parameter set
from a given survey, the Fisher matrix can also be used to
estimate the correlations between measured parameters. The
parameter correlation coefficient,
 \be
        \gamma_{ij} = \frac{\lgl \Delta \theta_i \Delta \theta_j \rgl}{
        \Delta \theta_i \Delta \theta_j} =
        \frac{F^{-1}_{ij}}{\sqrt{F^{-1}_{ii} F^{-1}_{jj}}},
 \ee
 shows the degree of degeneracy between two parameters and allows
 one to identify the largest eigenvalue of the parameter
 covariance matrix. This is of
much use, as the degeneracy between parameters can be the major
limiting factor in parameter estimation. Knowing the principle
degeneracy in parameter space, one can choose to increase the set
of modes, preferentially adding information so as to decrease this
degeneracy using the GOMA data compression analysis discussed
in Taylor et al. (2001).

\section{Cosmological Random Fields}
\label{cos_ran_fields}

The design of galaxy redshift surveys has been investigated before
by Heavens and Taylor (1997) who discussed the
optimisation problem, and by Tegmark (1997) who calculated the
uncertainty on parameters in linear theory, ignoring the effect on
redshift-space distortions. Taylor and Watts (2001) generalised
this to the nonlinear regime, and included both redshift
distortions and nonlinear bias. Here we summarise these results
for the linear analysis, assuming a general stochastic biasing
model~(Dekel and Lahav 1999) to relate the matter density field to
the galaxy field.

\subsection{Galaxy density and radial velocity fields}

 We shall assume for our analysis that the relevant field is the
 linear galaxy redshift-space density perturbation in the
 plane-parallel approximation. We write
 \be
        \delta^s_g(\k) =
        D(k \sigma_v \mu)[(b_L + \mu^2 f)\delta_m(\k) + \epsilon]
 \ee
 where $f\equiv d\ln \delta/ d \ln a \approx
 \Omega_{m}^{0.6}$~(Peebles 1980)
 is the growth index of density perturbations,
  $b_L$ is a linear bias parameter, $\delta_m(\k)$ is the
 linear matter density field, $\mu = \rhatb . \khat$ is the cosine
 angle between the wave-vectors of the density field and the
 observers line of sight and $\epsilon$ is a stochastic bias term. We
 assume 
 $\langle \epsilon \rangle = 0$ and $\langle \epsilon^2 \rangle =\sigma_\epsilon^2$.
 We have also included a nonlinear redshift distortion term
 \be
    D(k \sigma_v \mu) = (1+ k^2\sigma_v^2 \mu^2/2)^{-1/2}
 \ee
 where $\sigma_v$ is the pairwise radial velocity dispersion.

 We define a second bias parameter as the ratio of power in the
 galaxy distribution $P_{gg}(k)$, to that in the matter fields,$P_{mm}(k)$ :
 \be
    b  = \frac{P_{gg}(k)}{P_{mm}(k)} =
    b^2_L + \frac{\sigma^2_\epsilon}{P_{mm}(k)}.
\label{b(k)}
 \ee
 We may also introduce a galaxy correlation coefficient, $r_g$,
 defined by
 \be
    r_g = \frac{P_{gm}(k)}{\sqrt{P_{gg}(k)P_{mm}(k)}} = \frac{b_L}{b}
 \ee
 (Dekel and Lahav 1999) which will allow us to change between the two bias parameters.
 Although $b_L$ is more physical, it is $b$ which is more commonly
 measured.

 We choose our second field to be the radial gradient of the radial peculiar velocity field,
  defined in Fourier
 space as
 \be
    u^{\prime} (\k)=\frac{\partial}{\partial r}\rhatb . \vb(\k) = - \mu H f(\Omega_m)
    \delta_m(\k)
 \ee
where $H$ is the Hubble parameter. 
The modes of the radial velocity are uncorrelated with the modes of
the density field. However the Fourier space radial gradient of the
radial velocity \textit{is} correlated and will yield a cross power
spectrum.

 The auto- and cross-power spectra of these two fields are
 \ba
    P^s_{gg}(\k) &=& D^2(k \sigma_v \mu) ( 1+ 2 \mu^2 r_g \beta + \mu^4 \beta^2)
     b^2 P_{mm}(k) , \\
      P_{u^{\prime}u^{\prime}}(\k) &=& \mu^4 H^2 \beta^2 b^2 P_{mm}(k), \\
    P^s_{gu^{\prime}}(\k) &=& - \mu^2 H \beta  D(k \sigma_v \mu) (r_g + \mu^2 \beta) b^2
    P_{mm}(k),
    \label{power}
 \ea
 where
 \be
        \beta \equiv \frac{f(\Omega_m)}{b}
 \ee
 is the linear redshift distortion parameter.

 The three spectra to be measured are the redshift-space galaxy power
spectrum, $P^s_{gg}(\k)$, the radial velocity gradient power spectra,
$P_{u^{\prime}u^{\prime}}(\k)$, and the cross-spectra of these fields,
$P^s_{gu^{\prime}}(\k)$.

 The noise terms associated with these fields are
 \ba
    N_{gg}(r) &=& \frac{1}{n_g(r)}, \\
    N_{u^{\prime}u^{\prime}}(r) &=& \mu^{2} k^{2} \frac{\sigma_{DI}^2(r)}{n_v(r)},\\
    N_{gu^{\prime}}(r) &=& 0 ,
 \label{xcov}
 \ea
 where $n_g(r)$ and $n_v(r)$ are the number densities of the
 galaxy and velocity surveys, respectively. The factors of $\mu^2$ and
 $k^2$ are due to our use of the radial velocity gradient. We quantify
 the number densities  in
 Sections 4 and 7. The noise term for radial velocities arises from the intrinsic
 uncertainty on galaxy positions due to the dispersion in the
 $D_n$-$\sigma$ relation for ellipticals. This can be approximated
 by
 \be
    \sigma_{DI}(r) = \sigma_0 H r e^{r/r_{\rm err}},
 \ee
 where $\sigma_0$ is the fractional error on the distance
 indicator. We allow the
 fractional distance error to blow up beyond
 some fiducial distance, $r_{\rm err}$. We choose a conservative
 $r_{\rm err}$ of $135 \hMpc$ (which is 90 $\%$ of the 6dF median depth),
 but allowing this to decrease to as low as $50\%$ of 
the median depth had
 little effect on our results.

\subsection{The matter power spectrum and cosmological parameters}
 Throughout we define the matter power spectrum by
 \be
    P_{mm}(k) = Q^2 k^n T^2(k;\Gamma, \omega_b),
 \ee
 where $Q$ is the amplitude of matter perturbations, $n$ is the
 primordial spectral index of perturbations and $T(k)$ is the
 matter transfer function. The dependent parameters of the transfer
 function are the so-called ``shape parameter'',
 \be
    \Gamma = \Omega_m h
 \ee
 which stretches the scale of the transfer function,
 and
 \be
    \omega_b = \Omega_b h,
 \ee
 where $\Omega_b$ is the density parameter of baryonic matter.
 Here we use the form of the transfer function given by Eisenstein
 and Hu (1998).

 The other parameters which will be of interest to us are the
 amplitude of galaxy clustering and the amplitude of the velocity field.
 As we shall assume linear theory throughout, these are
 respectively
 \be
    A_g = b Q
 \ee
 where $b$ is the linear galaxy bias parameter, and
 \be
    A_v = f(\Omega_m) Q.
 \ee
 Also of interest is the mass-galaxy correlation
 coefficient $r_g$, and 
  as we shall assume the galaxy survey is in
 redshift-space, the final parameter is the redshift-space
 distortion parameter, $\beta$.

 For the purposes of this paper we shall assume the primordial
 spectrum is scale-invariant so that $n=1$. Hence the parameters
 we shall investigate are the six non-degenerate parameter set
 $(A_g,A_v,\Gamma,\omega_b,\beta,r_g)$. The model we shall use has
 fiducial values of $(Q,h,\Omega_m,\omega_b,b,r_g)=(5\times 10^{-5},
 0.65,0.3,0.025,1,1)$ giving the measurable parameters the values
 $(A_g,A_v,\Gamma,\omega_b,\beta,r_g)=(5 \times 10^{-5},0.486 \times 5 \times 10^{-5},
 0.195,0.025,0.486,1)$.

\section{The galaxy redshift survey}

\subsection{Survey parameters}

A galaxy redshift survey can be parameterised by the fractional
sky coverage, $f_\sky$, and the mean radial density of galaxies in
the sample, which we parameterise by 
 \be
    n_g(r) = N_g \frac{a }{r_*^3 \Gamma[3/a]} e^{-(r/r_*)^a}
\label{numden}
 \ee
 where $r_*$ is a convenient survey depth scale, $N_g$ is
the total number of galaxies and typically $a=1.5$.

Rather than use $r_*$ we use the median depth
of the survey ($r_{\rm m} \approx 1.5r_*$) and the number
of galaxies available in the survey roughly scales as
 \be
    N_g \approx 4. \times 10^{-2} \alpha r_{\rm m}^{3.125},
 \ee
from fitting the near infrared differential number counts compiled in
Kochaneck et al. (2000). 

 The
sampling fraction, $\alpha$, is simply the fraction of sources
expected at any redshift which are actually observed. We shall
assume that space is Euclidean. The three survey parameters are
then $f_\sky$, $r_m$ and $\alpha$.

\subsection{Optimisation of a Galaxy Redshift Survey}
\label{optgrs}

The problem of optimisation of a galaxy redshift survey has been previously
considered by Heavens and Taylor (1997), who showed that the
optimisation for a fixed timescale reduces to a one-parameter
optimisation problem. We first need to choose a parameter for which to
optimise. Here we choose the amplitude of galaxy perturbations,
$A_g$, where the conditional error on a measurement is given by
 \be
    \frac{\Delta A_g}{A_g} =
    \left[\frac{1}{2}\int \! \frac{d^3 k}{(2 \pi)^3} \,
    V_\eff^g(k,\mu)\right]^{-1/2}.
 \ee

\begin{figure}
\centering
\begin{picture}(200,230)
\includegraphics{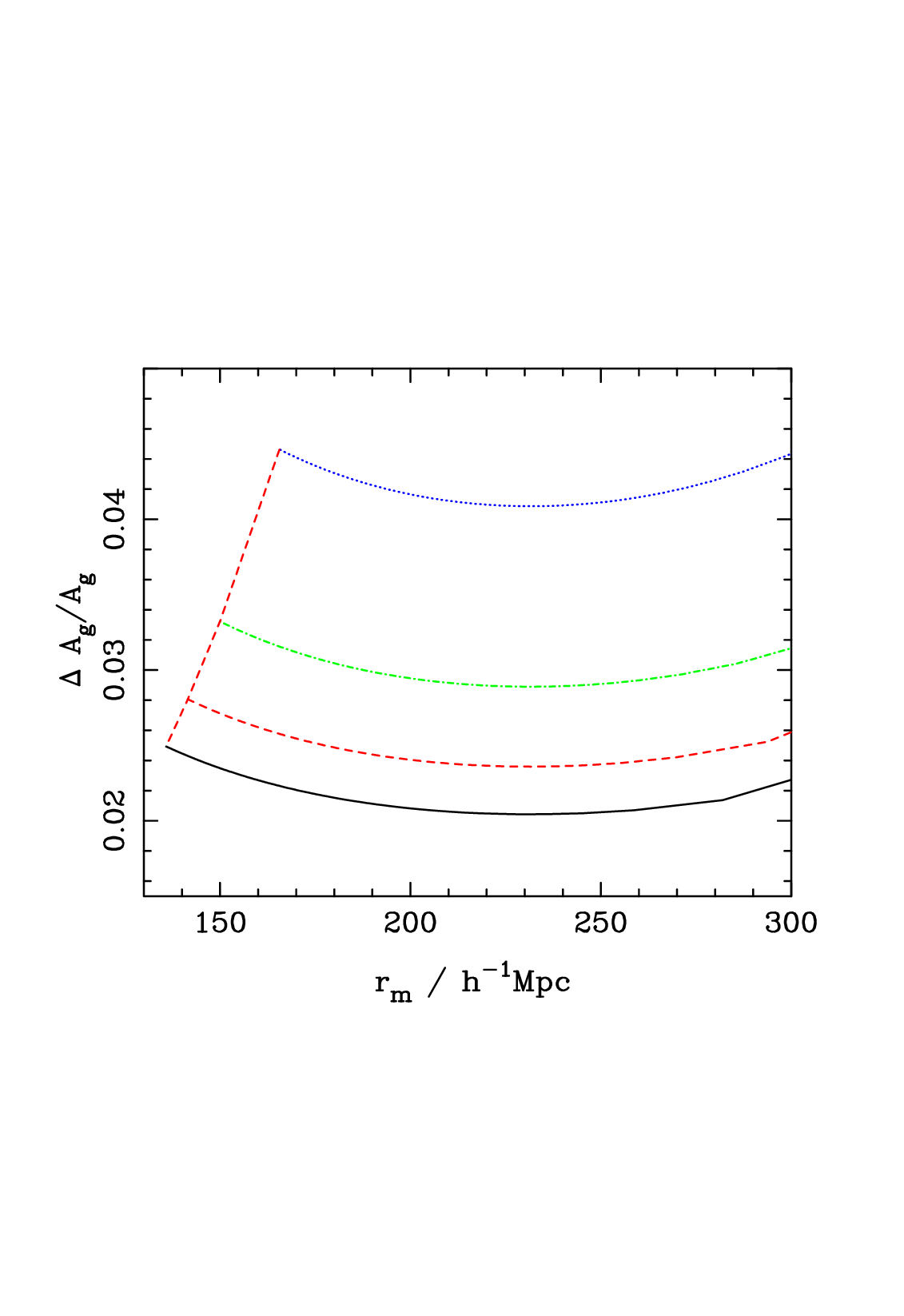}
\end{picture}
\caption{Optimisation of the 6dF redshift survey. Here we plot the fractional error on
a conditional measurement of the amplitude of galaxy clustering,
$A_g$, as a function of survey median depth, $r_{m}$. The lines are for constant survey
timescales with sky coverage of unity (bottom), three quarters 
(second from bottom), a hemisphere (third from bottom) and a quarter 
(top). }
\label{pub9}
\end{figure}

We now assume the timescale for the redshift survey is a constant and scales
as
 \be
    t =t_0  \alpha f_\sky (r_m/h^{-1} {\rm Mpc})^7 ,
 \ee
 where $t_0$ is some fixed timescale. We take as our baseline
 a 2 year survey where $t/t_o \approx 7 \times 10^{17}$.
 The strongest constraint then comes from the survey depth.
 The uncertainty on parameters scales as $\Delta \theta \propto
f_\sky^{-1/2}$, while the time scales as $t \propto f_\sky$,
indicating, as is well known, that as large a solid angle as
possible should be chosen. With this time constraint the survey
median depth is
\be
    r_m = 150 (\alpha f_\sky)^{-1/7} h^{-1}{\rm Mpc}.
\label{timeconstr}
\ee
 This states that the depth of the survey is increased at the
 expense of sparse sampling or reduced sky coverage.
 We can now minimise the uncertainty on $A_g$.

 Figure \ref{pub9} shows the results of minimising the fractional
 error on the galaxy clustering amplitude as a function of survey depth, for a fixed timescale. Each
 of the lines is for a different sky coverage. The approximate relation we use
 between limiting K band magnitude and median depth is
\be
 K_{limit} \approx 5\log \left( \frac{r_{m}}{h^{-1}{\rm Mpc}} \right) +1.865.
\ee
The sampling fraction for each value of $r_{m}$ may be calculated from
 equation~(\ref{timeconstr}). The dashed line in Figure~\ref{pub9} shows
 the physical limit $\alpha = 1$.  
 The optimal survey is a shallow all-sky survey with depth
 $r_{\rm m}=140 h^{-1}{\rm Mpc}$. For a hemisphere this increases slightly to
 $r_{\rm m}=150 h^{-1}{\rm Mpc}$, while the optimal value of $\alpha \approx
 0.7$ is fairly
 insensitive to either $f_{sky}$ or $r_m$. The
 increase in the uncertainty in the fractional error in $A_g$ with
 increased depth is due to the
 decrease in the total number of galaxies in the survey. At the
 extreme of very large $r$ the fractional uncertainty in $A_g$ tends
 towards $A_g \sim r^{6.5}$.

\begin{figure}
\centering
\begin{picture}(200,230)
\includegraphics{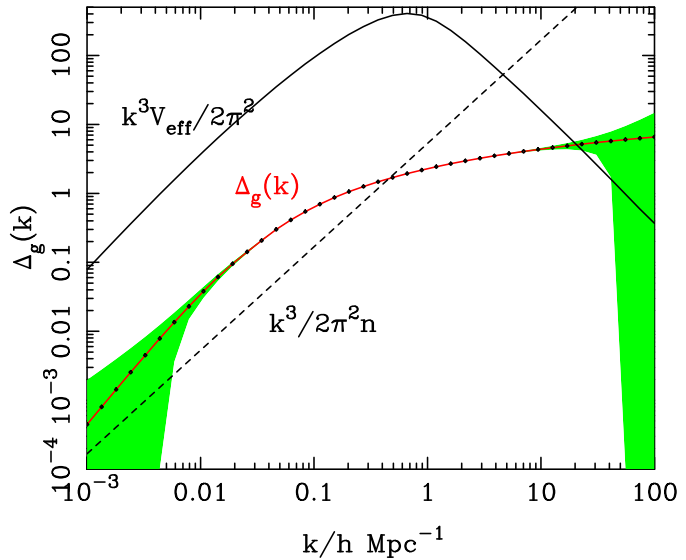}
\end{picture}
\caption{The linear galaxy power spectrum with fiducial LCDM
parameters (solid light line). Also plotted are the expected band
averaged error on measured power pass-bands (grey area), and the
effective survey volume $V^g_{\rm eff}(k)$ (dark solid line) and
the shot-noise per mode (dashed line), for  the 6dF redshift survey. We parameterise the
6dF survey by its sky
coverage, depth and total number of galaxies.} \label{pub15}
\end{figure}

Figure \ref{pub15} shows the linear galaxy power spectrum using
the optimal redshift survey parameters for the 6dF, along with the expected
conditional error bars in band averaged pass-bands of width
$\Delta \ln k =0.5$. We have also plotted the effective volume,
$V^g_{\rm eff}(k)$, which indicates where the information is
maximised in the survey. Finally we have also plotted the noise
per mode, $k^{3}/2 \pi^{2}\bar{n}$, where $\bar{n}$ is the average
density of the survey. For the rest of the analysis we use $K_{limit}
=12$, $f_{sky} = 0.5$, $r_m = 150 h^{-1}$Mpc and $\alpha = 0.7$.

\section{Parameter forecasts for the 6\lowercase{d}F redshift survey}

In linear theory a galaxy redshift survey based on a CDM model
contains three free parameters: the amplitude of the galaxy power
spectrum, $A_g = b Q$, the scale of the break, parameterised by
$\Gamma=\Omega_m h$ and the redshift distortion parameter,
$\beta=\Omega^{0.6}/b$. Other parameters that could be considered
are the small-scale pairwise velocity dispersion, $\sigma_v$, or the baryonic
density parameter, $\Omega_b$, which can introduce oscillations in
the the matter power spectrum. Other possible parameters include
the spectral index $n$~(Tegmark 1997) or the correlation between
luminous and dark matter $r_{g}$. We consider the estimation of
five parameters $A_{g}$, $\Gamma$, $\beta$, $\Omega_{b}$ and $r_{g}$.

\begin{figure}
\centering
\begin{picture}(200,230)
\includegraphics{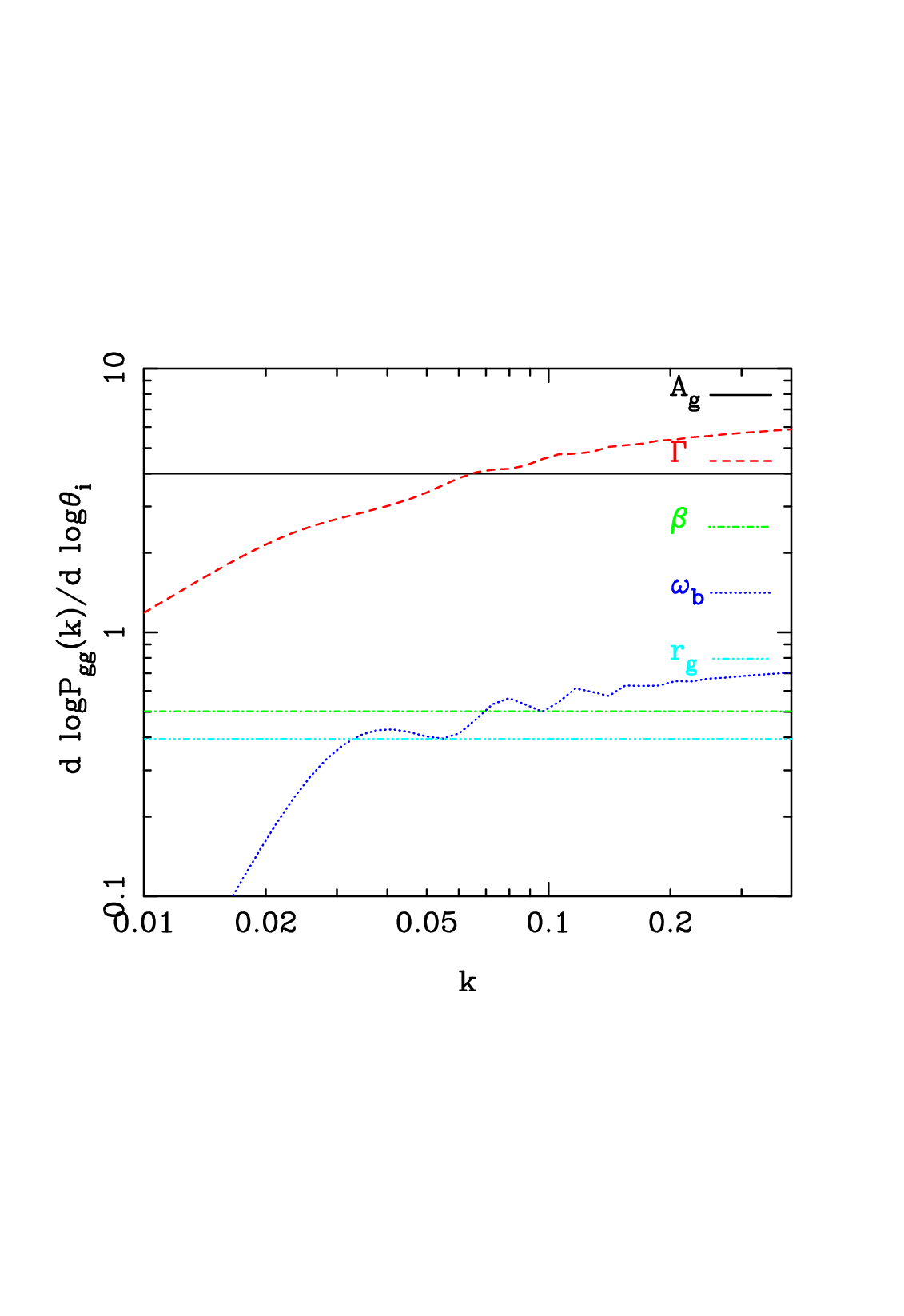}
\end{picture}
\caption{Derivatives of the matter power spectrum } \label{pub16}
\end{figure}

\subsection{The information content of the redshift-space galaxy power spectrum}
\label{info}

The information content of a galaxy redshift power spectrum can be
examined per Fourier mode by plotting the derivatives of the
log-power with respect to the parameters, $d \ln P^s_{gg}(k)/d
\theta$. In geometric terms this derivative can be thought of as a
vector in Hilbert space, and the Fisher matrix as the dot product
of such vectors. Hence the more similar two curves are, the more
``parallel" they are and so the greater the degeneracy between the
parameters. In the extreme that two curves are identical the
Fisher matrix becomes singular and the uncertainty becomes
infinite.

Figure~\ref{pub16} shows that the derivatives for the five
parameters we can extract from linear theory, $\thetab =
(A_{g},\Gamma, \beta, \omega_b,r_g)$. The $A_{g}$-derivative is a constant
across all modes, as expected. Similarly both $r_{g}$ and $\beta$
are almost constants, as they appear in the normalisation of the
galaxy power spectrum. Hence we can expect some degeneracy between
these three parameters.

The final two parameter derivatives, $\Gamma$ and $\omega_b$ are
also similar. This is because $\Gamma$ parameterises the break in
the CDM power spectrum, between retarded modes inside the horizon
at matter-radiation equality and growing modes outside, while
$\omega_b$ parameterises the effect of baryonic damping of
perturbations inside the horizon when the baryonic fluid is
oscillating. Hence both parameters are related to two physically
different effects that produce similar damping of the matter power
spectra. An effective shape parameter can be defined by~(Peacock and
Dodds 1994)
 \be
    \Gamma_{\rm eff} = \Gamma e^{-2 \Omega_b h}
 \ee
 which crudely models both effects but explicitly shows the
 degeneracy. On scales larger than the break, the effects of the
 baryons dies away quicker, and so on the largest scales
 the power spectra contains more
 information on $\Gamma$ than on $\omega_b$.


\subsection{The amplitude of galaxy clustering and redshift
distortion parameter}
\label{betasec}

In the limit of negligible shot noise, it is possible to calculate
analytically the two-parameter Fisher matrix for $A_g$ and
$\beta$. The details of this are shown in Appendix A. From
equation~(\ref{betacorr}) the correlation
coefficient between the two parameters is  $\gamma_{A_g
\beta}=-0.78$, which is very close the value found by Taylor
et al (2001) for the Point Source Redshift Survey (PSCz). The strong
anti-correlation is due to the similar effects of the parameters on
the power spectrum, as they both appear in the amplitude of the
redshift-space power. 
From equation~(\ref{errormatrix}) we can read off the fractional
errors: $\Delta A_g/A_g=0.01$, and $\Delta \beta/\beta =0.057$ for
current typical survey sizes at the limit of linear theory $k \approx
0.2 h {\rm Mpc}^{-1}$.

 For a single parameter estimate of $\beta$, again in the absence of
 shot noise we find;
 \be
    \frac{\Delta \beta}{\beta}
    = \sqrt{\frac{2\pi^2}{V k^3}}
    \left(2+ \frac{1}{1+\beta} -
    \frac{3 \tan^{-1}(\sqrt{\beta})}{\sqrt{\beta}}\right)^{-1/2},
    \label{betaerr}
 \ee
 where $V$ is the survey volume, and $k$ is the maximum wavenumber
 used to estimate $\beta$. For $\beta=0.5$, and $k=0.2 \Mpc$ this
 gives a fractional uncertainty of
 \be
    \frac{\Delta \beta }{ \beta} =
    \left( \frac{V}{4.5\times 10^{4} [\hMpc]^3} \right)^{-1/2}.
 \ee
 Hence, even for a well-sampled galaxy redshift survey, the
effective volume has to be large to beat down the sampling noise
on a determination of $\beta$. To achieve a $1\%$ error, assuming
all other parameters are known, requires a survey with volume
$V\approx 4.5 \times 10^8 [ \hMpc]^3$, over four times larger than
the 2dF, 6dF or SDSS redshift surveys. Hence to achieve high accuracy
on cosmological parameters in the linear regime requires an order
of magnitude increase in survey size. However even for these
surveys, with volumes of order $V \approx 10^8 [\hMpc]^3$, we can
expect an accuracy of $\Delta \beta/\beta \approx 0.02$. An
alternative to going large is to extend the analysis into the
nonlinear regime, as outlined by Taylor \& Watts (2001). Finally,
the uncertainty can be reduced by combining data sets, as
investigated here.

\begin{figure}
\centering
\begin{picture}(200,400)
\includegraphics{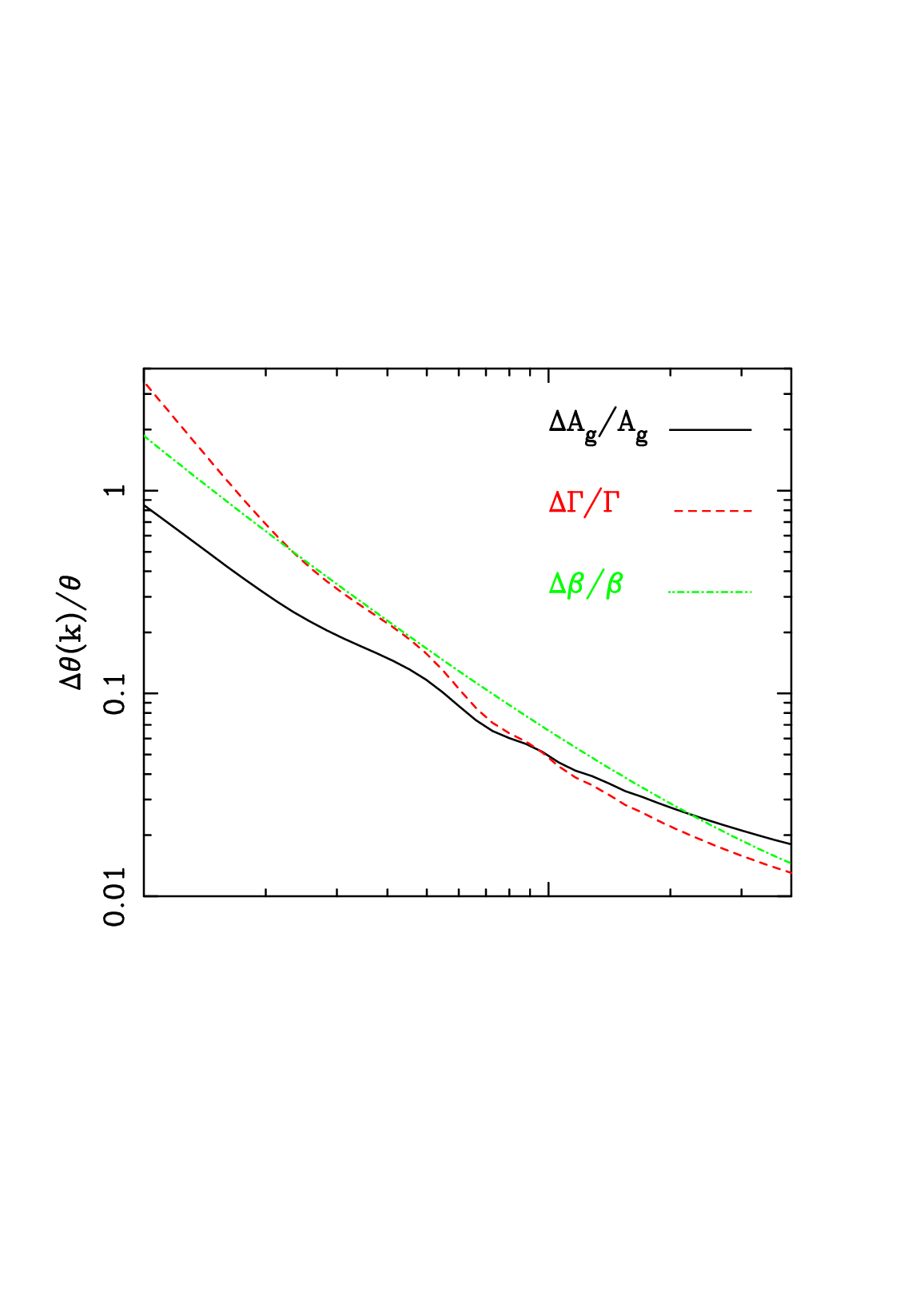} \includegraphics{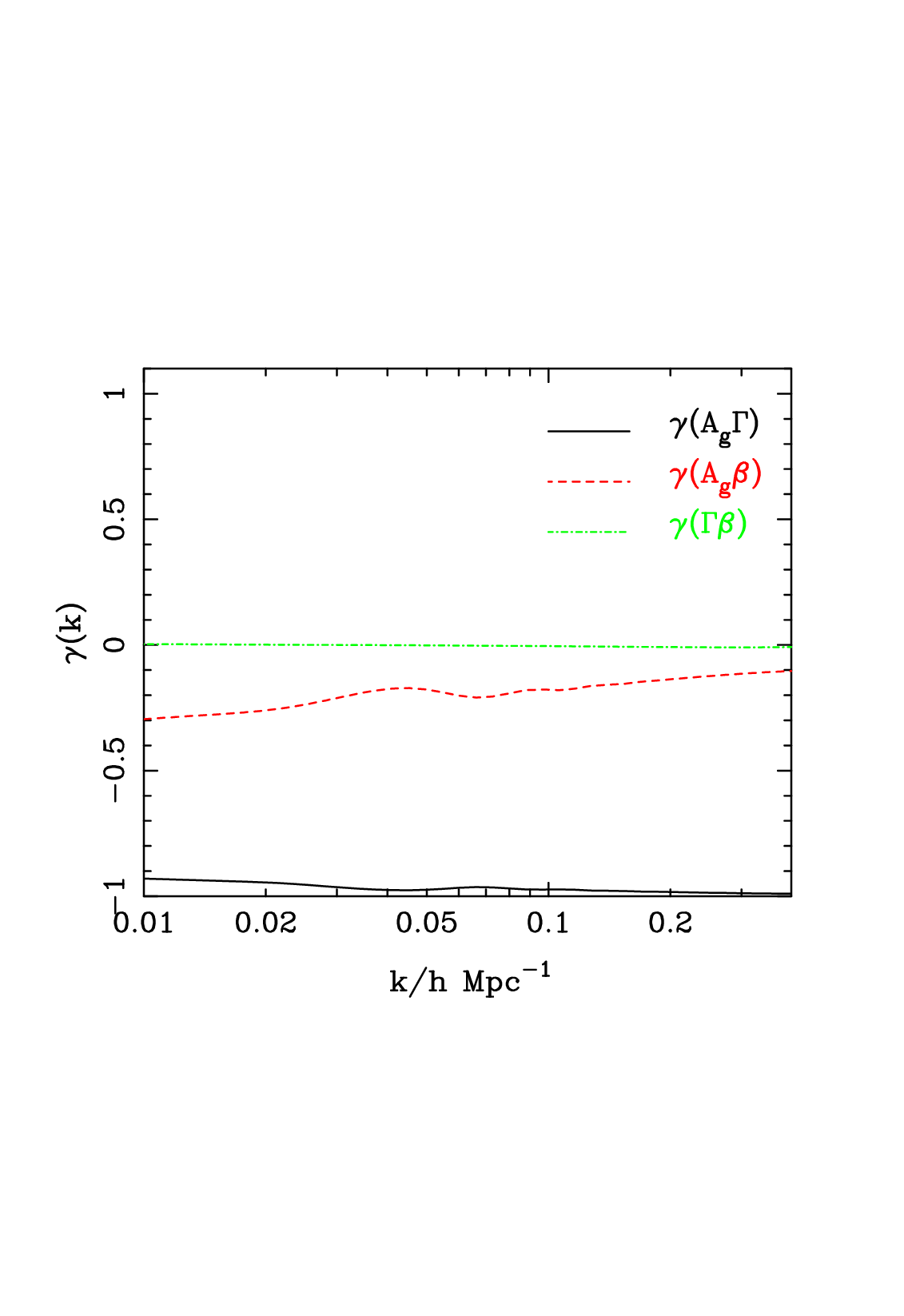}
\end{picture}
\caption{{\bf Upper plot:} The expected fractional uncertainty on
$A_g$, $\Gamma$ and $\beta$ from the 6dF redshift survey. {\bf Lower plot:} The
correlations between $A_g$, $\Gamma$ and $\beta$. We model the survey
as described in the text.} \label{pub1}
\end{figure}

\subsection{Three parameter set: $A_{g}$, $\Gamma$ and $\beta$}

 Introducing extra correlated parameters into the
analysis causes uncertainties to increase, and adds extra
complexity to the analysis. Hence we first consider a reduced
3-parameter set consisting of $A_{g}$, $\Gamma$ and $\beta$, and then
see how adding new parameters changes our results.

Figure \ref{pub1} shows the expected fractional marginalised
uncertainties from the 6dF redshift survey plotted as a function of maximum
wavenumber analyzed. The scale of interest for linear analysis is
around $k\approx$ $0.2 \Mpch$. At this wavelength the best
estimated parameter is $\Gamma$ which is constrained to within
about $2\%$. The uncertainty on $A_g$ and $\beta$ is about $3\%$. For
$\beta$ this is close to our conditional estimate in Section \ref{betasec}.

Figure \ref{pub1} also shows the correlation coefficient for these
parameters. The strongest correlation is between the amplitude of
galaxy clustering, $A_g$, and the shape parameter, $\Gamma$,
with $\gamma (A_g \Gamma) \approx -0.95$. This arises because a
change in amplitude can be mimiced by a shift in scale. The least
correlated parameters are $\Gamma$ and $\beta$.

Intermediate is the correlation between $A_g$ and $\beta$. This is
of particular interest as their combination gives (Taylor et al
2001)
 \be
    A_g \beta = Q f(\Omega_m) \approx Q \Omega_m^{0.6},
 \ee
 and so provides an independent estimate of the amplitude of
 matter perturbations. The correlation of these parameters can be
 calculated in the limit of negligible shot-noise, and
 ignoring other parameters is given by equation (\ref{betacorr})
 in Appendix A.
 For $\beta=0.5$ we find $\gamma_{A_g \beta}=-0.78$, very close to
 the value we find for a two parameter analysis.

 As these parameters are anti-correlated, the combination $A_g
 \beta$ has a smaller formal error than either $A_g$
 or $\beta$. This arises because the combination $A_g \beta$
 marginalises over the $A_g$-$\beta$ parameter space along the
 longest line of the degeneracy, so that the marginalised
 combination has a smaller error than its
 components. This effect is shown later, in Section~\ref{comb} figure~\ref{3ellipse}.

\begin{figure}
\centering
\begin{picture}(200,400)
\includegraphics{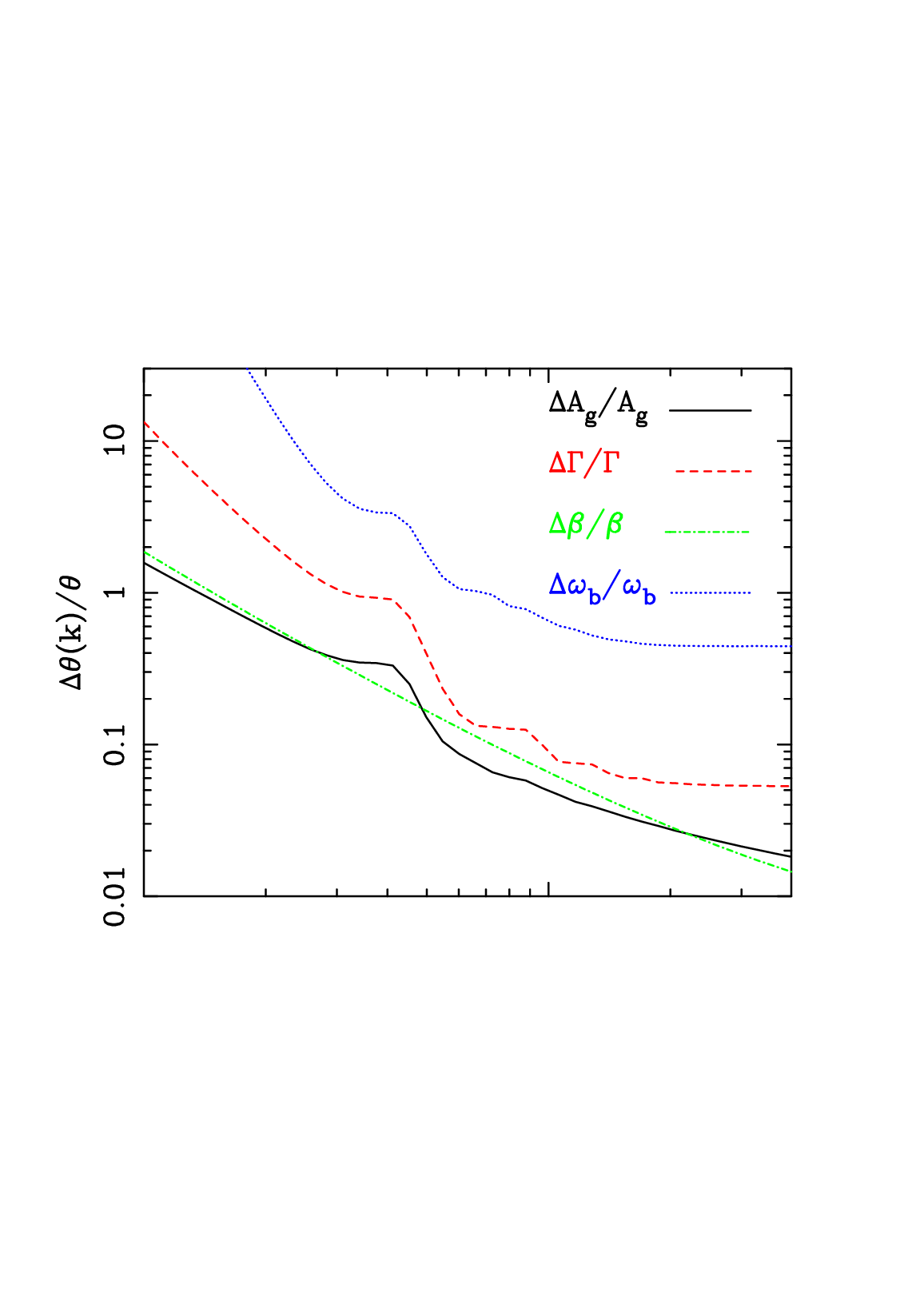}\includegraphics{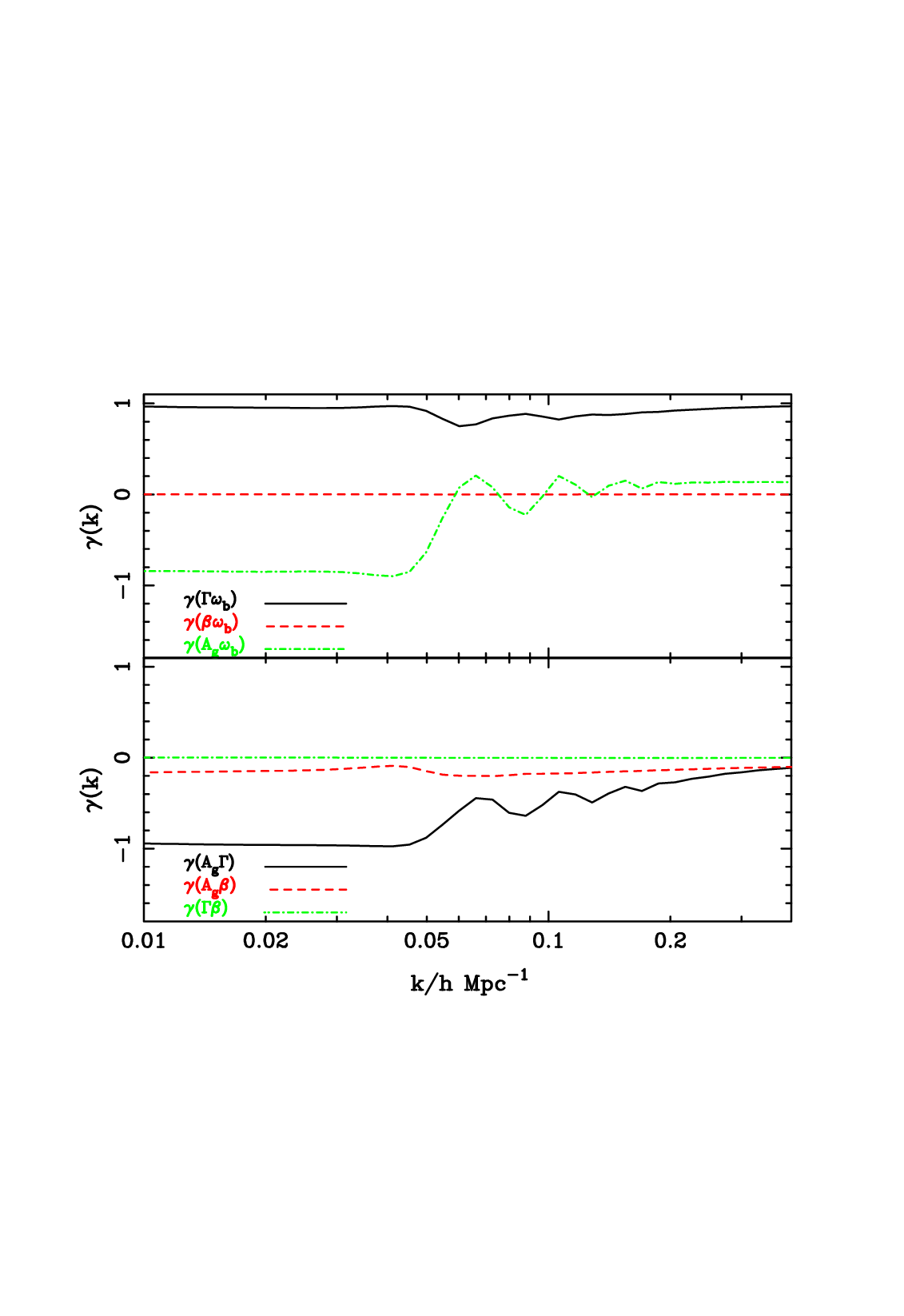} 
\end{picture}
\caption{{\bf Upper plot:} Predicted errors for a four parameter
analysis involving $\omega_{b}$. {\bf Lower plot:} Correlations in
a four parameter space with $\omega_{b}$ } \label{pub3}
\end{figure}

\subsection{Four parameter set: $A_{g}$, $\Gamma$, $\beta$ and $\omega_{b}$}
\label{galfour}

Introducing a fourth parameter, $\omega_b$, has little effect on
the uncertainties on $A_g$ and $\beta$ on scales $k>0.05
\Mpch$, as shown in Figure \ref{pub3}. Figure \ref{pub3} shows
that this is due to the weak correlation between $\omega_b$ and
these parameters. However the strong correlation between
$\omega_b$ and $\Gamma$, $\gamma (\Gamma \omega_b) \approx 0.9$,
which we expect from our discussion in Section \ref{info},
degrades the uncertainty on $\Gamma$. Disappointingly the
uncertainty on a measurement of $\omega_b$ is large, $\Delta
\omega_b/\omega_b \approx 0.45$, which reflects its weak effect on
the matter power spectrum. As with all the parameters, the main
way to improve on sensitivity is with a larger redshift survey.

 Since $\omega_b$ and $\Gamma$ are correlated would we do any
 better by estimating the baryon fraction,
 \be
    f_b = \frac{\omega_b}{\Gamma} = \frac{\Omega_b}{\Omega_m},
 \ee
 rather than simply $\omega_b$? Unfortunately not. Although these
 parameters are positively correlated, and so the marginalised
 combination will project out along the longest axis of the
 $\omega_b$-$\Gamma$ error ellipse, making the fractional error on
 $f_b$ smaller, other correlations, particularly between $\omega_b$ and
 $A_g$, prevent any improvement. We have re-analysed the
 survey with $f_b$ replacing $\omega_b$ throughout and find
 negligible change in our results. We conclude that the main
 limiting factor here is the survey volume.

 In order to reduce the
 error on $\omega_b$ to the $10\%$ level we predict one would
 require a survey some 10 times larger.

\begin{figure}
\centering
\begin{picture}(200,400)
\includegraphics{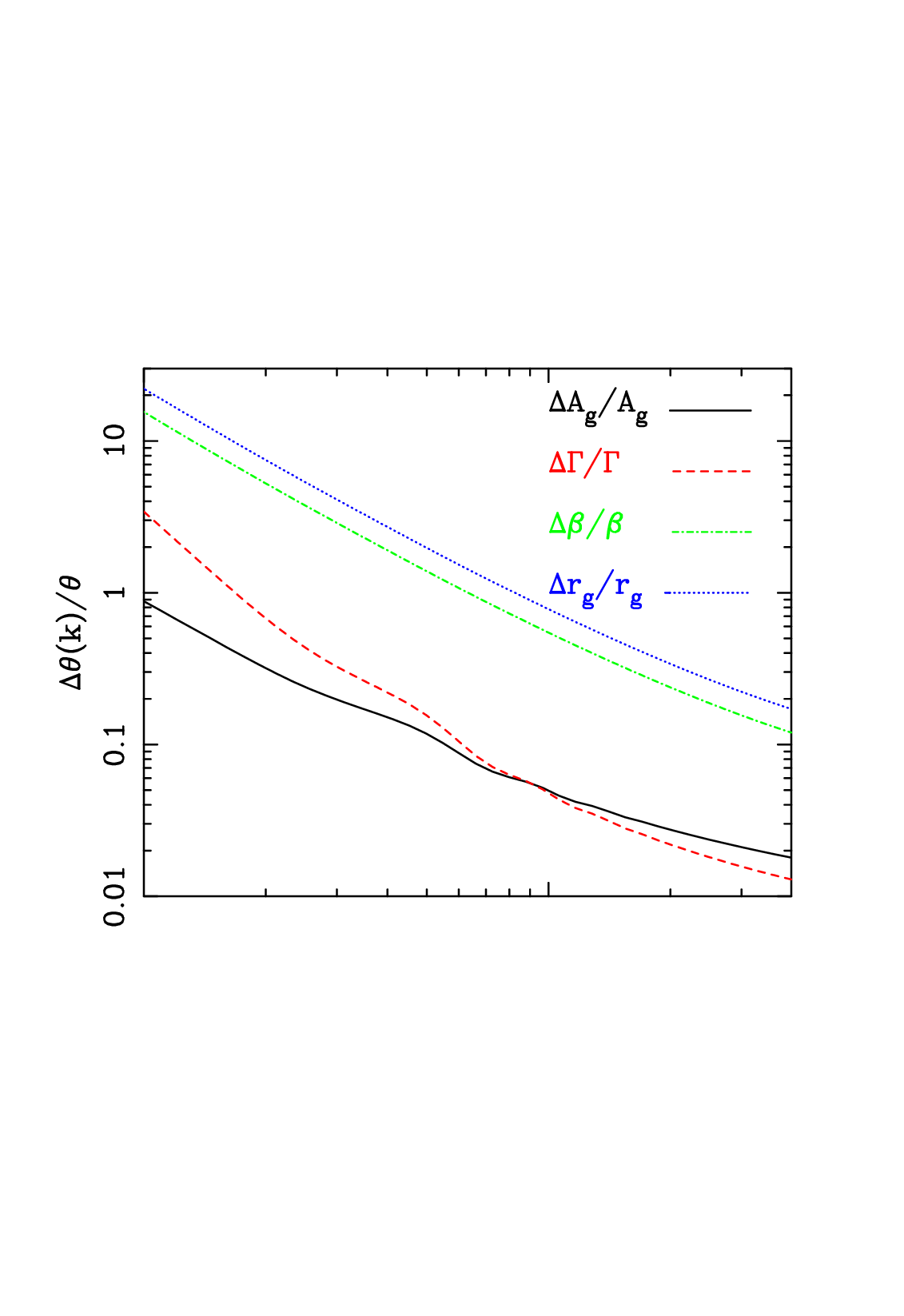} \includegraphics{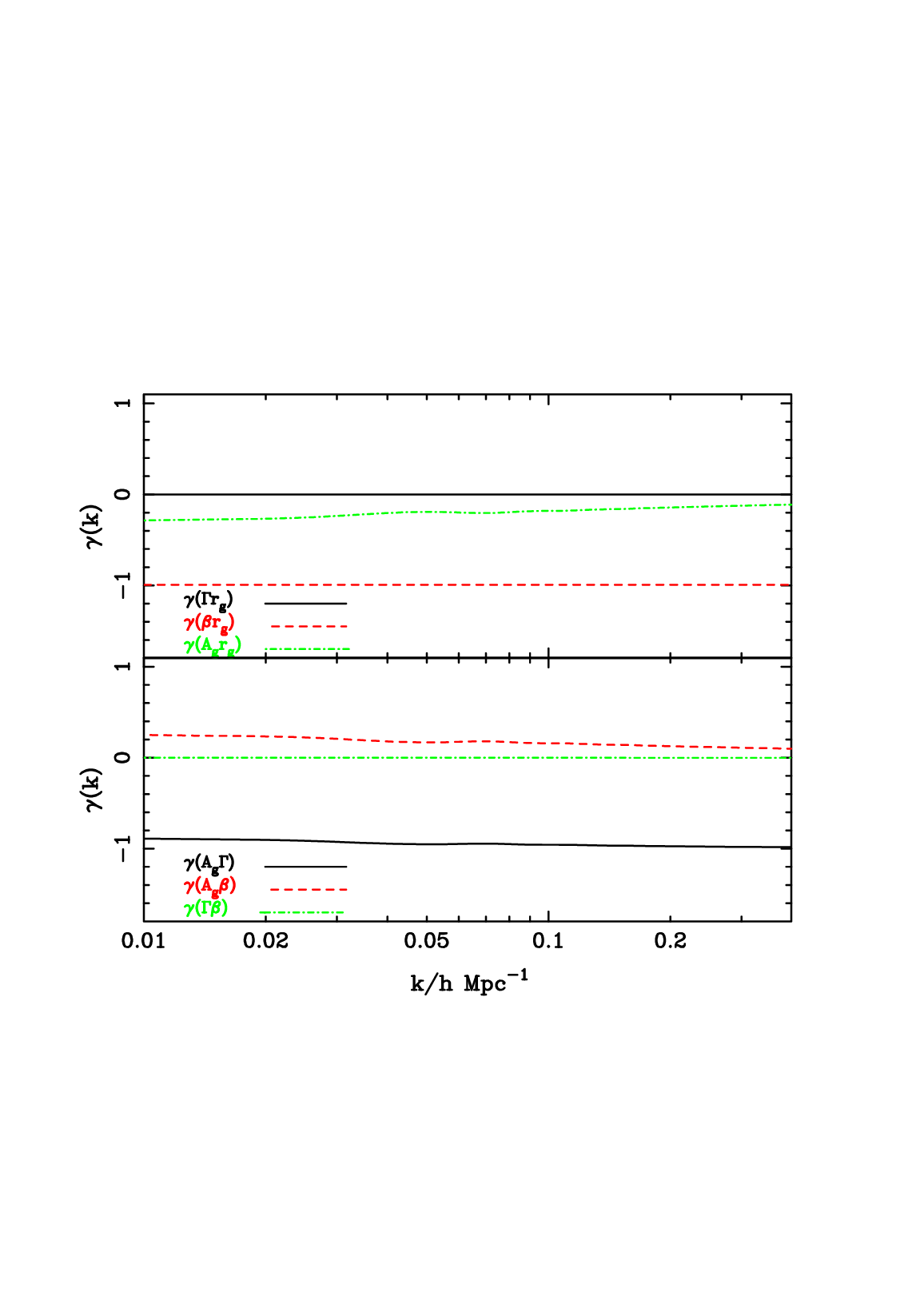}
\end{picture}
\caption{{\bf Lower plot:} Uncertainties for a four parameter
analysis invloving $r_{g}$. {\bf Lower plot:} The correlations
between $A_{g}$, $\Gamma$, $\beta$ and $r_{g}$ } \label{pub5}
\end{figure}

\subsection{Four parameter set: $A_{g}$, $\Gamma$, $\beta$ and $r_{g}$}
 Instead of $\omega_b$, we could have chosen $r_g$ as our fourth
 parameter. Figure \ref{pub5} shows that this time $A_g$ and
 $\Gamma$ are unaffected by this new parameter, but the
 uncertainty on $\beta$ is dramatically increased, with both the fractional
 errors on $\beta$ and $r_g$ around $35\%$. Again this can
 be traced to the expected degeneracy between $r_g$ and $\beta$
 discussed in Section \ref{info}, and shown in Figure \ref{pub5}, where
 $\gamma_{\beta r_g} \approx -1$. Interestingly the correlation
 between $A_g$ and $\beta$ has now become positive, so that the
 error on their combination will be increased.

\section{Galaxy Velocity Survey}
The application of Fisher information methods to galaxy velocity
surveys has so far been limited. Here we derived for the first
time the Fisher matrix from parameters from such a radial velocity
survey, and use it to find the optimal survey parameters, estimate
the conditional errors on a measurement of the velocity power
spectrum and estimate the marginalised errors on cosmological
parameters. In Section \ref{comb} we combine the radial velocity
and galaxy redshift surveys.

\begin{figure}
\centering
\begin{picture}(200,230)
\includegraphics{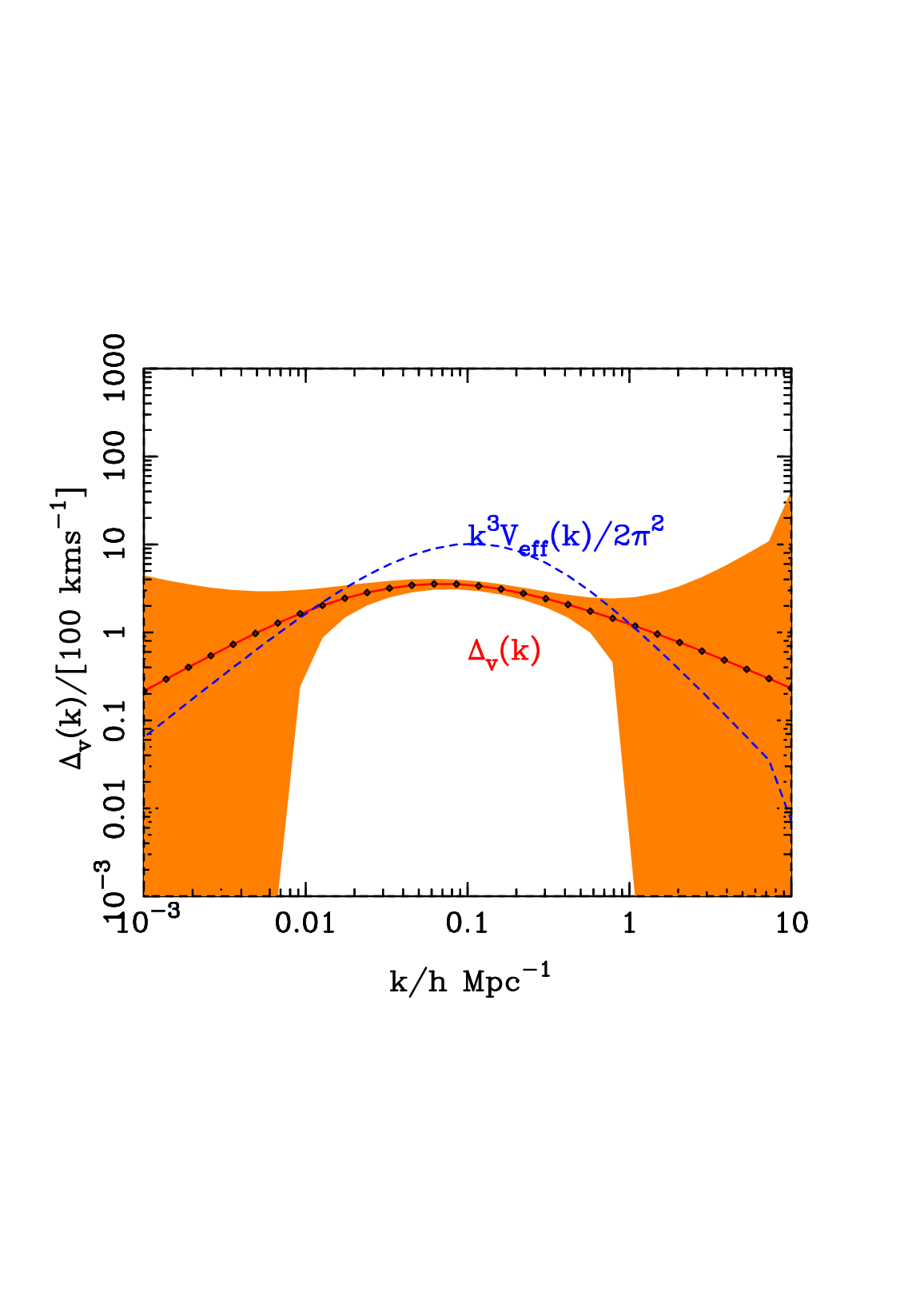}
\end{picture}
\caption{The peculiar velocity power spectrum, $\Delta_v(k)=\sqrt{k^3 P_v(k)/2
\pi^2}$ with predicted noise and effective volume for the 6dF peculiar
velocity survey. The
velocity survey is parameterised by $f_{sky}$, cut-off depth, $\alpha_v$ and
the accuracy of the distance indicator.}
\label{pub25}
\end{figure}

 Figure \ref{pub25} shows the 3-D velocity power spectrum for our
 fiducial LCDM cosmology, and the effective survey volume, $k^3
 V^v_\eff(k)/2 \pi^2$, for a model 6dF velocity survey.

\section{Velocity survey optimisation}

\subsection{Velocity survey parameters}

The velocity survey can be parameterised by the fractional sky
coverage, $f_\sky$ and the mean radial density of galaxies in the
sample, which is parameterised as for the galaxies (equation~\ref{numden}). The early type galaxies for the velocity survey will be a
subset of this group and so for analysis of the velocity survey
the number count needs to be multiplied by a factor
of $0.3$ and the analysis limited to $r_m$ which is the
cut-off in the 6dF velocity survey. We shall again assume a Euclidean
universe for simplicity, and as we are working at low redshift.
The survey can then be parameterised by $f_\sky$, $r_m$, $\alpha_v$
and $\sigma_0$, the expected fractional error in the distance indicator.

The angular part of the effective volume for the velocity survey
can be integrated analytically and yields
 \be
    V^u_{\rm eff}(k) = 2 \pi f_{\rm sky} \int_0^\infty \! dr r^2 \,
    {\Theta}\left[\frac{n_u(r)P_{uu}(k)}{\sigma_{DI}^2(r)}\right]
 \ee
 where the function ${\Theta}(z)$ is defined in Appendix A,
 equation (\ref{defA}). Figure \ref{pub25} shows the dimensionless
 effective survey area. The peak of the effective volume again
 indicates where the information content is maximised.

\subsection{Band-averaged power}

As well as showing the effective volume Figure~\ref{pub25} shows the error bars expected
 for a measurement of band-averaged velocity power in a series of
 pass-bands, using equation (\ref{bandpow}). The power spectrum
 plotted is the three dimensional velocity power spectrum but the
 error bars reflect the fact that this will be derived using only radial
 velocity data. The pass bands are of
 width $\Delta \ln k =0.5$ and so this plot may be directly compared
 with the corresponding band averaged power for the redshift survey in
 Figure~\ref{pub15}. The velocity power spectrum has much larger error
 bars which is a reflection of the smaller size of the velocity data set.

\subsection{Optimisation of a velocity survey}

As with the redshift survey we wish to know whether the velocity survey can be
optimised in the sense of maximising the information content for
the minimum observing  time. To optimise the survey, we choose the
amplitude of the power spectrum as a fiducial optimisation
parameter, where
 \be
    \frac{\Delta A_v}{A_v} = \left[\int \!
    \frac{d \ln k}{4 \pi^2} \, k^3 V_\eff^v(k)\right]^{-1/2},
 \ee
 and we consider the case of equal-time surveys.

The timescale for a velocity survey scales as the total number of
sources in the survey, $N \sim r^3$, the reciprocal of the
effective flux of the sources, $S\sim r^{-2}$, the distance error
per source, $\sigma_{DI}$, the fraction of the sky surveyed,
$f_\sky$, and the fraction of the sources sampled, $\alpha$. The
timesale for a velocity survey then scales as
 \be
    t \approx  \alpha_v f_\sky \left( \frac{\sigma_0}{ {\rm km}s^{-1}}
\right)^{-2} \left( \frac{r_m}{h^{-1}{\rm Mpc}} \right)^7 t_0
 \ee
 where $t_0$ is a constant timescale. For a survey of a few
 years we find the ratio $t/t_0 \approx 2 \times 10^{13}$.

We can now minimise the fractional uncertainty on the amplitude of
the velocity power spectrum,  $A_v$, in survey-space with the
equal-time constraint and the restrictions $f_\sky$, $\alpha_v \le
1$. In our model, the overall uncertainty on parameters scales as
$\Delta \theta \propto f_\sky^{-1/2}$, so all our results and
timescales can be scaled by the fraction of sky surveyed. In the
following analysis we fix the value of $f_\sky$ and optimise in
terms of the remaining survey parameters. In addition the
uncertainty in source distance per source, $\sigma_0$, and the
source sampling rate, $\alpha_v$, always appear in the ratio
$\sigma_0^2/\alpha_v$, further reducing the number of free survey
parameters we need to consider. Finally, we can relate the
remaining parameter, $r_m$ to the other parameters by
 \be
    r_m \approx 150 \left[ \frac{1} {\alpha_v f_\sky} \left(
 \frac{\sigma_0} {{\rm km}s^{-1}} \right)^2 \right]^{1/7} h^{-1}Mpc.
 \ee
 Hence we only need to optimise the survey with respect to the
degenerate parameter
 \be
    \nu=\frac{1}{\sqrt{\alpha_v}} \left( \frac{\sigma_0}{{\rm km}s^{-1}}
\right).
 \ee
 In practice there is a lower limit to the accuracy of the
distance indicator. The current best accuracies from the 
$D_{n}$-$\sigma$ relation provide distance estimates with a minimum scatter of $\sigma_{{\rm min}}\approx 0.1 {\rm km}s^{-1}$ e.g.~(Jorgensen
et al. 1993; Gregg 1995),
and so we shall optimise the survey design in terms of the new
parameter
 \be
    \eta=\frac {1} {\sqrt{\alpha_v}} \left( \frac{\sigma_0 +
\sigma_{{\rm min}} }{{\rm km}s^{-1}} \right)
 \ee
 which better reflects the survey's limitations.

\begin{figure}
\centering
\begin{picture}(200,230)
\includegraphics{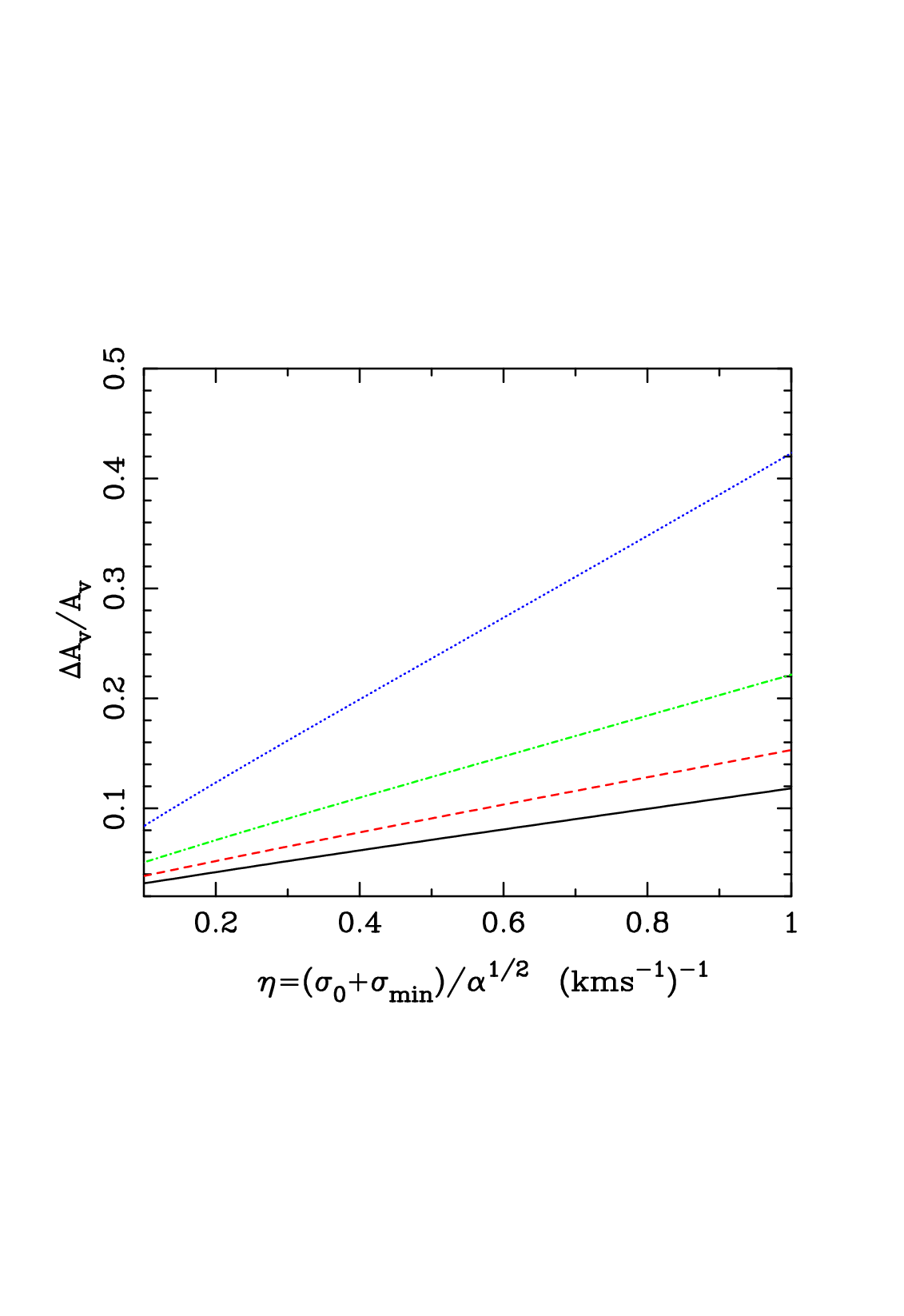}
\end{picture}
\caption{The normalised fractional uncertainty on the amplitude of
the velocity power spectrum for the degenerate parameter
$\eta=\frac{(\sigma_0 +\sigma_{{\rm min}})} {{\rm kms^{-1}}}\frac{1}{\sqrt{\alpha_v}}$, for
equal-time surveys. The four lines correspond to different sky
fractions (see Figure~\ref{pub9}).} \label{pub23}
\end{figure}

 Figure \ref{pub23} shows the fractional uncertainty on a conditional
 measurement of $A_v$ for equal-time surveys as a function of
 $\eta$ and where the different curves correspond to different
 sky fractions, $f_\sky$. The lowest, solid line is for an all-sky
 survey, with higher lines reduced by one quarter.
 The expected 6dF velocity survey lies somewhere between the top two
 lines depending on dust extinction from the plane of the galaxy.

The other parameter that this analysis will depend on is the upper
wavenumber used in the analysis, $k_{\rm max}$. Linear theory is
reasonably robust up to $k=0.2\Mpc$. In practice it was found that
the optimal survey parameters changed little with changes in
$k_{max}$, so that our results with $k_{max}=0.2 \Mpch$ are quite
general.

Our results show that the optimal strategy is to be as accurate as
possible in the distance determinations at the expense of going
deep. In fact the absolute optimal strategy is found to be
impossible with current distance indicator limits -- if the
distances could be better constrained then we could do even
better. The reason for this is that by $k \approx 0.2 \Mpch$ the
power spectrum is already becoming dominated by shot noise. If
there were more sources (and so less shot noise) this would have
the same effect as increasing $\alpha$ in the degenerate parameter
$\eta$ and so the optimal value for $\sigma_{0}$ would also
increase. The bottom line is that the survey should be as well
sampled and as accurate as possible before going deep.

For the rest of this analysis we assume $\sigma = 0.1 kms^{-1}$,
$\alpha_v =0.2$, $f_{sky} = 0.5$ and $\eta = 0.2$.

\section{Cosmological parameter forecasts from velocity
surveys}

\begin{figure}
\centering
\begin{picture}(200,400)
\includegraphics{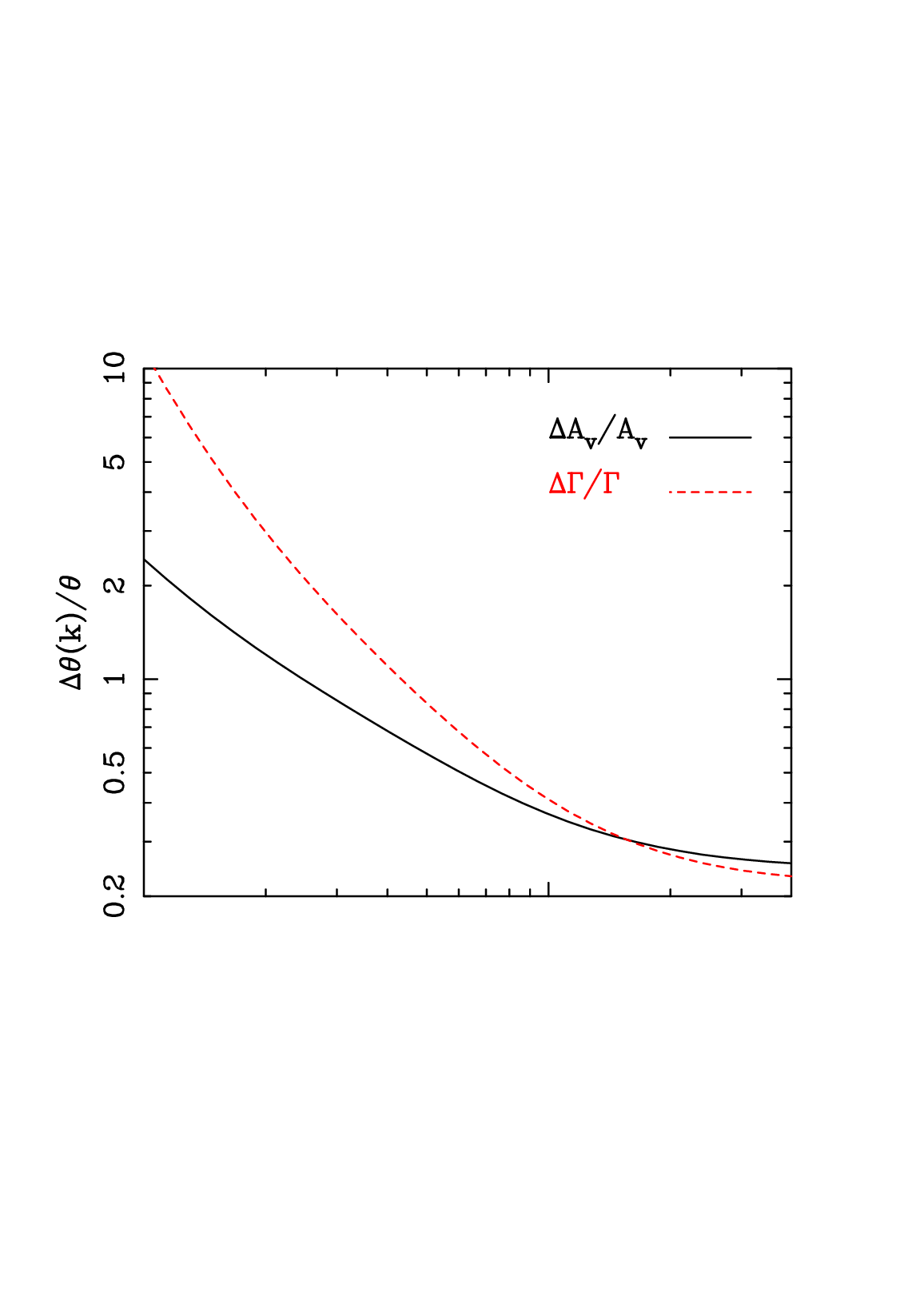}\includegraphics{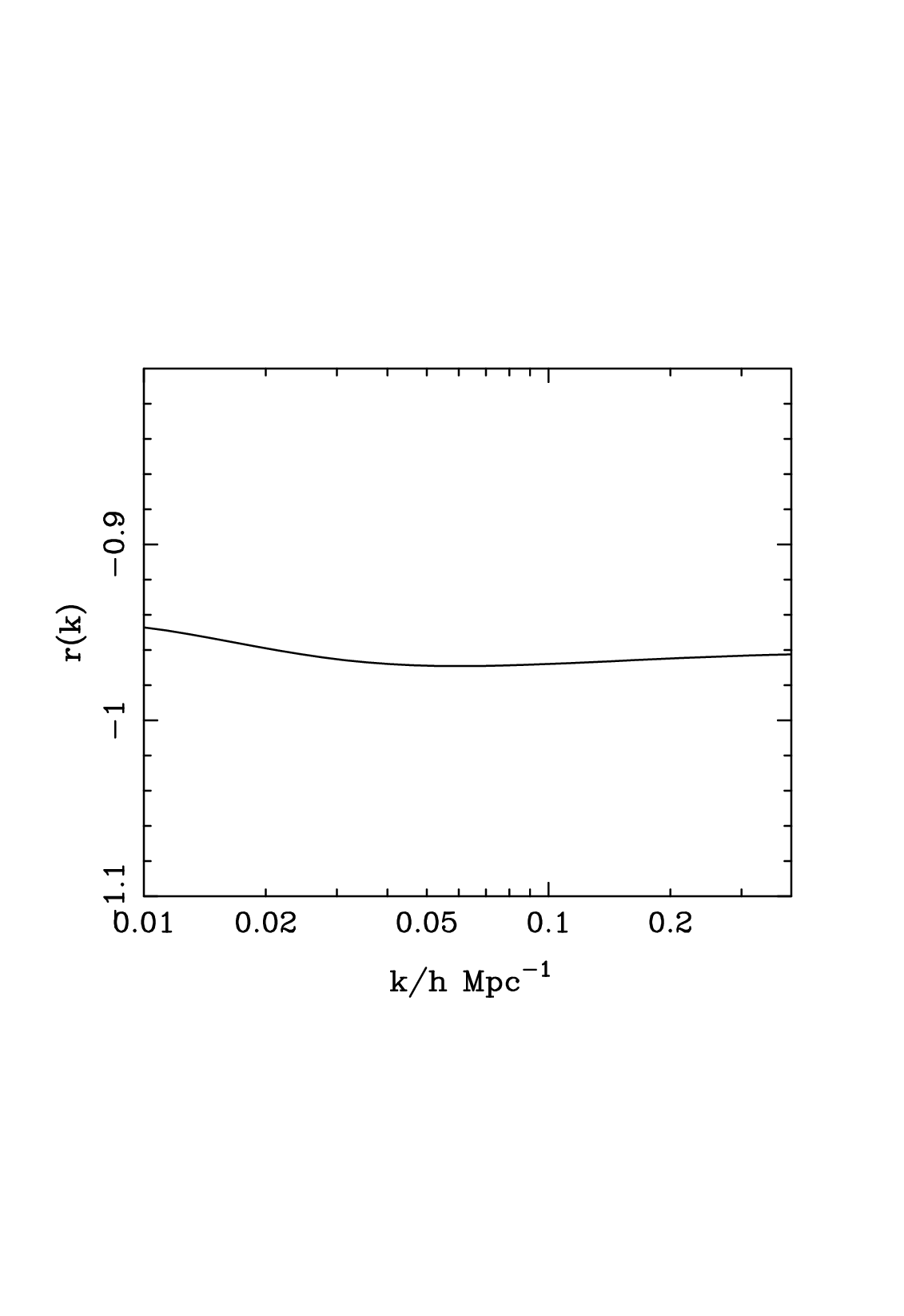}
\end{picture}
\caption{The expected fractional error (top) and correlation
coefficient (bottom) for the amplitude and shape parameter of the
6dF velocity power spectrum. The survey parameters are discussed in
the text.} \label{pub14}
\end{figure}

In the linear regime the properties of the velocity field can be
specified by just two parameters -- the amplitude of the velocity
power spectrum, $A_v$, and the shape parameter, $\Gamma$. Figure
\ref{pub14} shows the fractional error expected on $A_{v}$ and
$\Gamma$ for an optimal survey, as a function of maximum
wavenumber analyzed. If we truncate the analysis at $k=0.2\Mpc$,
the fractional uncertainty in both $A_{v}$ and $\Gamma$ is
expected to be $\approx 25 \%$, ignoring any systematic
uncertainties.

Figure \ref{pub14} also shows the parameter covariance coefficient
 \be
    \gamma_{A_v \Gamma}(k) = \frac{\lgl \Delta A_{v} \Delta \Gamma \rgl }
        {\Delta A_{v} \Delta \Gamma} \approx -0.8
 \ee
 as a function of wavenumber. It is because of this anticorrelation
 that a joint constaint of $A_v$ and $\Gamma$ is difficult.  
 If we assume a value for $\Gamma$ obtained from
 the redshift survey, Figure~\ref{pub23} shows that $A_v$ may be
 constained on its own to $\approx 5\%$.

\section{Combining data sets}
\label{comb}

 In this section we combine the information content of the radial
 velocity field and the galaxy redshift survey. We calculate the
 Fisher matrix using equation~(\ref{super}) which is
 written explicitly in Appendix B. As explained in
 Section~\ref{cos_ran_fields} we now work with the radial velocity
 gradient as this will yield a cross-power spectrum with the density
 field.

\subsection{Three parameter set}

\begin{figure}
\centering
\begin{picture}(200,400)
\includegraphics{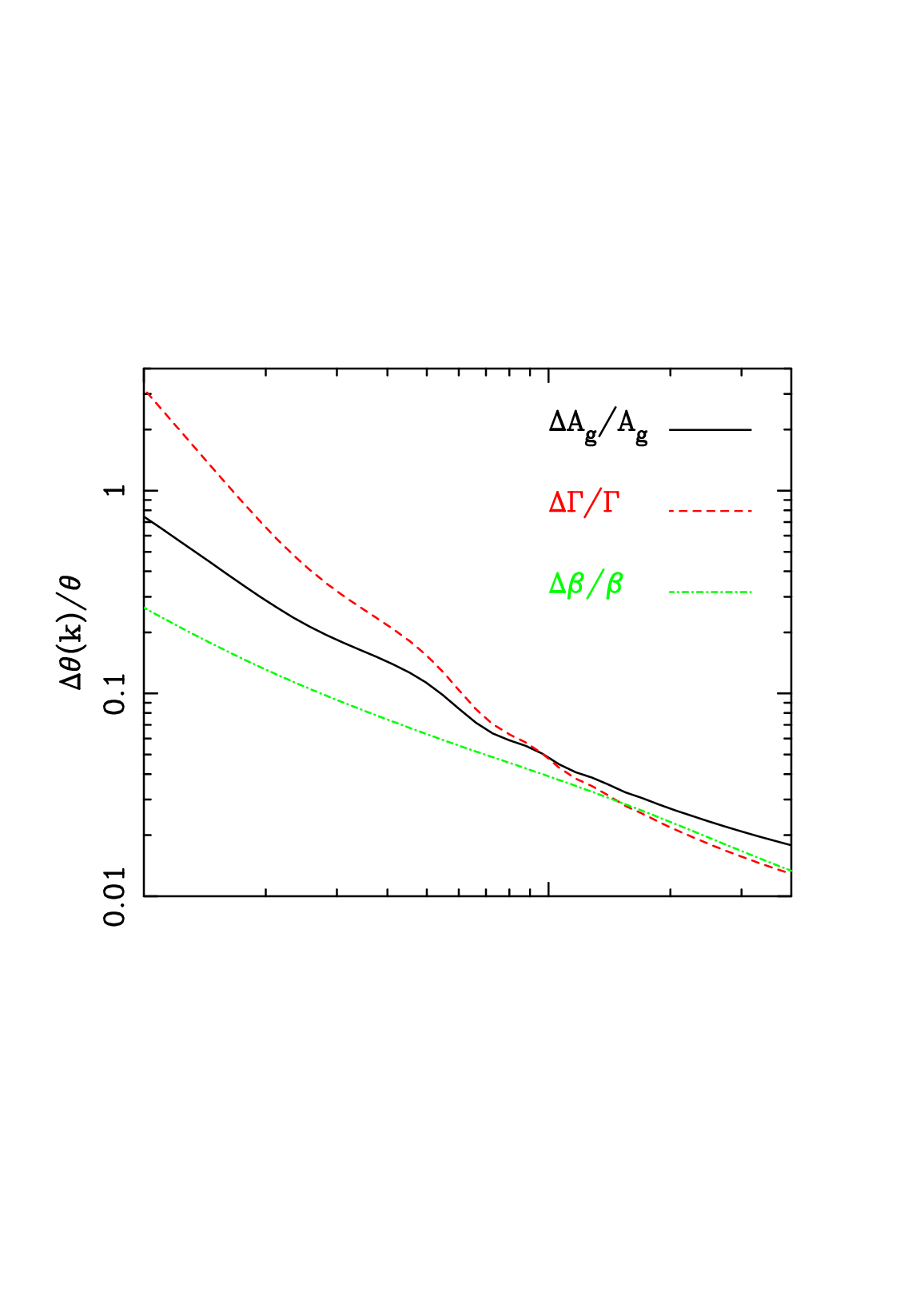}\includegraphics{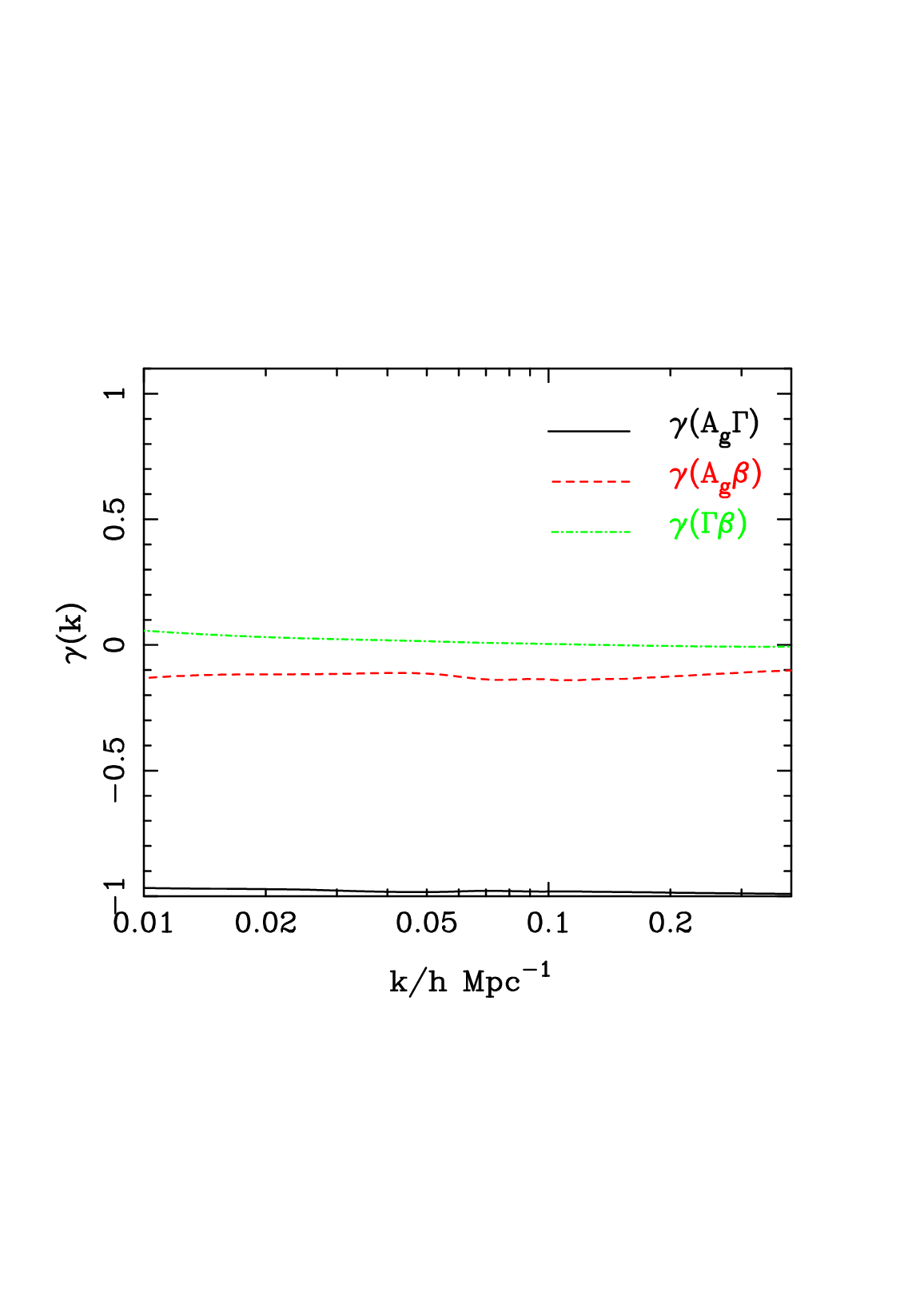}
\end{picture}
\caption{Supermatrix fractional errors and 
correlations between $A_{g}$, $\Gamma$ and $\beta$. These predictions are
for a combination of the 6dF redshift and peculiar velocity surveys.} \label{pub20}
\end{figure}

To see how combining redshift and velocity data affects parameter
estimation we begin with our three parameter set; $A_{g}$, $\Gamma$
and $\beta$. Figure \ref{pub20} shows the uncertainties and
correlations of this set. Comparing with just redshift
information, in Figure \ref{pub1} we see that the constraints on
$A_{g}$ and $\Gamma$ remain largely unchanged and there is a slight
improvement in the constraint on $\beta$. So for this three parameter set
there is relatively little improvement. This is because the inclusion
of the velocity data does nothing to break the main degeneracy between
$A_{g}$ and $\Gamma$. The other parameters were relatively uncorrelated. The
lower panel in Figure~\ref{pub20} shows that the correlations remain
unchanged from the redshifts-only analysis. 

\begin{figure}
\centering
\begin{picture}(200,270)
\includegraphics{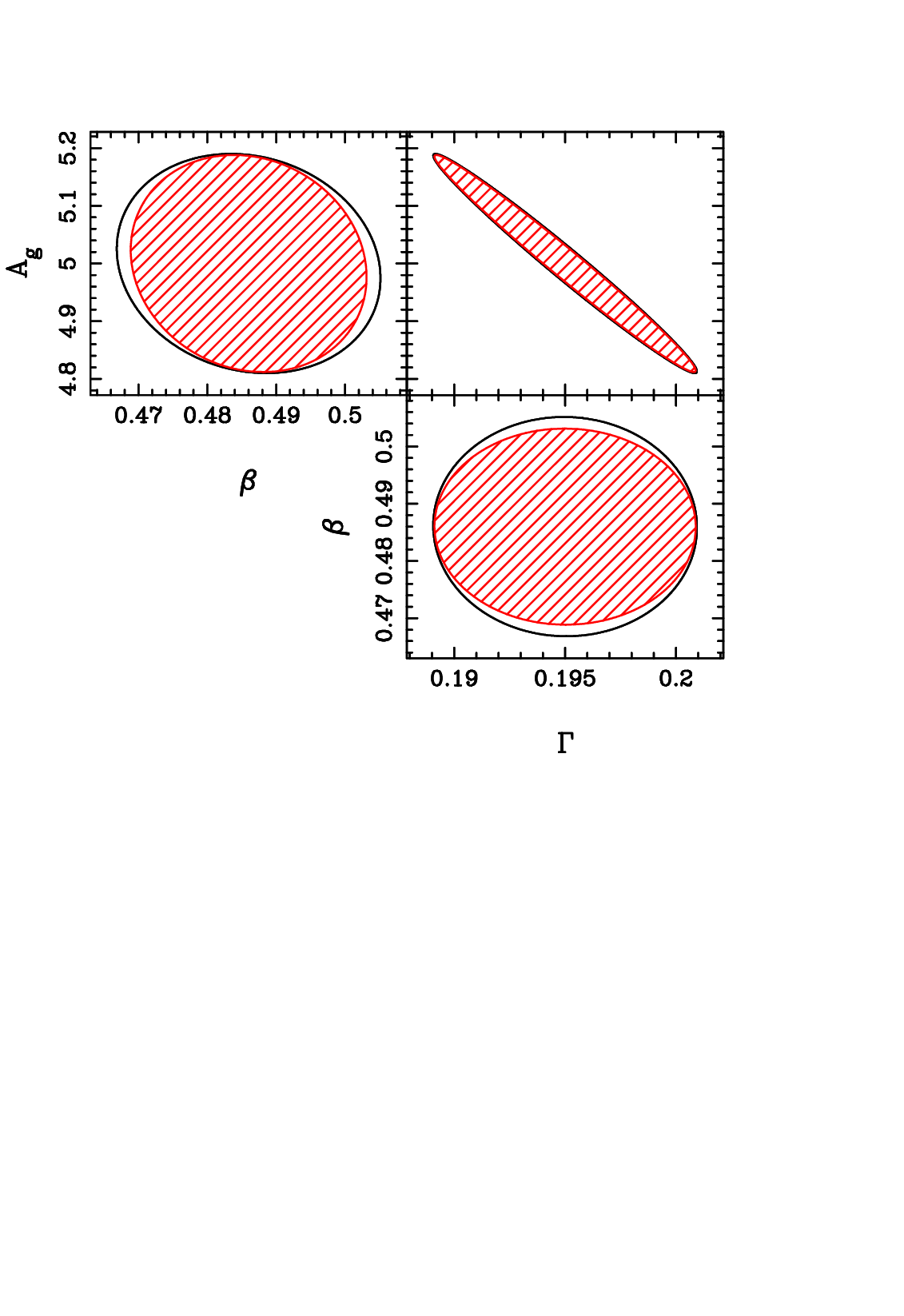}
\end{picture}
\caption{$1 \sigma$ contours for likelihood marginalised over one free
parameter. The outer contour is for the redshift survey only. Hatched regions indicate the
constraint from combining with the peculiar velocity survey.}
\label{3ellipse}
\end{figure}

Figure~\ref{3ellipse} shows the $ \Delta \ln {\rm(Likelihood)} =- \frac{1}{2}$ likelihood contours about
the fiducial maximum likelihood point in a two-parameter space. These
contours represent the likelihood marginalised in each
case over the one remaining parameter. The outer contour is that from
the redshift survey alone whereas the inner hatched region is the
constraint when velocity data is included. The maximum wavenumber
analysed here is $k=0.2 \hMpc$.

\subsection{Four parameter set}

As with the galaxy survey, we can add in a new parameter and study
its effect on a likelihood analysis. Adding in $\omega_b$ has little effect
on $A_{g}$ and $\beta$, but degrades $\Gamma$, as was the case for the
redshift survey alone. The baryonic density is not detected from the
combined survey anyway, nor is the baryon fraction. We shall not consider
this set further.

\begin{figure}
\centering
\begin{picture}(200,400)
\includegraphics{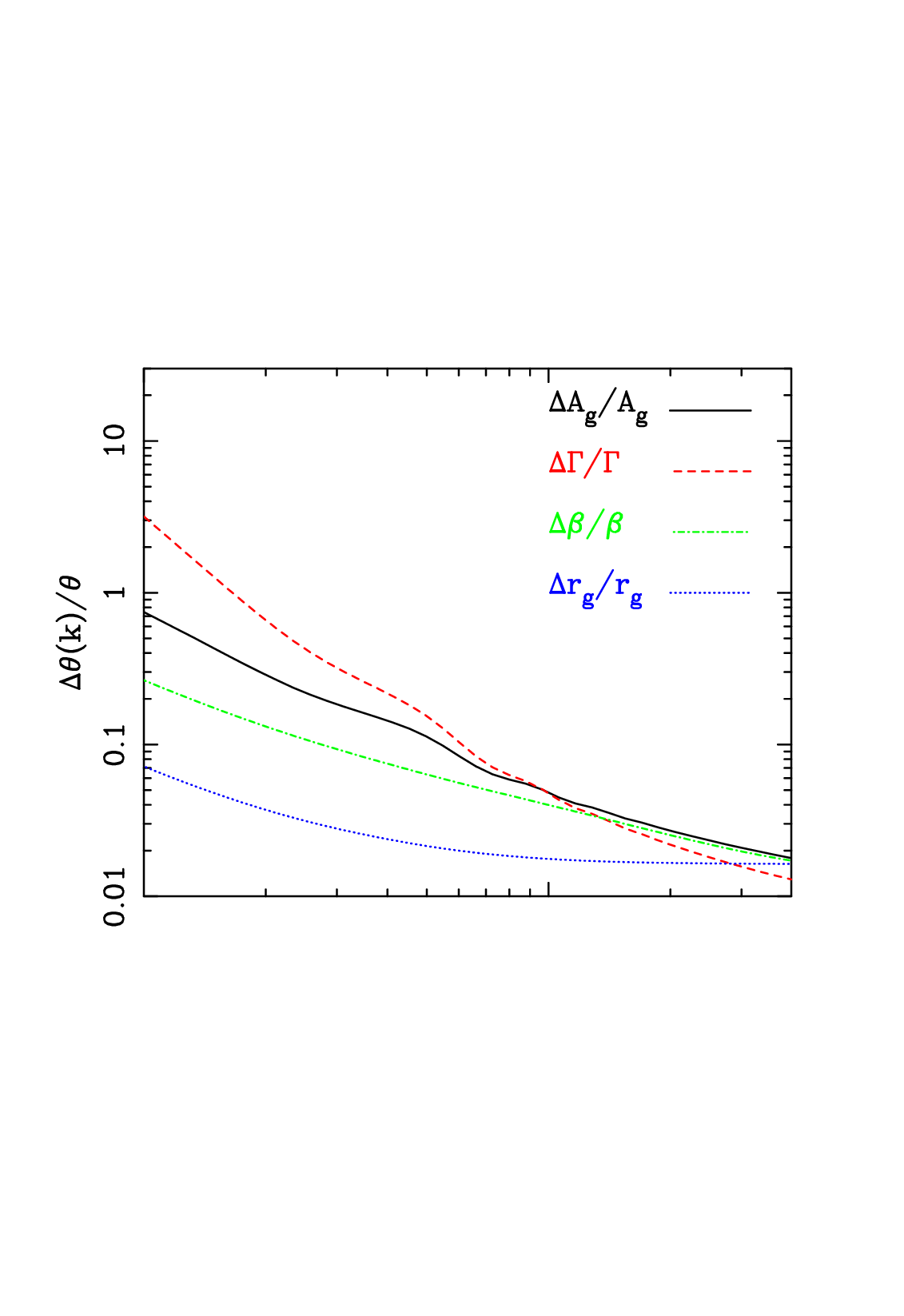}\includegraphics{figure12b.ps}
\end{picture}
\caption{Supermatrix fractional errors and
correlations between $A_{g}$, $\Gamma$, $\beta$ and $r_{g}$.}
\label{pub22}
\end{figure}

 Figure \ref{pub22} shows the marginalised uncertainties of a
 four parameter set including the
 galaxy-mass correlation parameter, $r_g$. Comparing with Figure~\ref{pub20}
 we see that $A_g$ and $\Gamma$ are unchanged by the inclusion of the
 new parameter, since these are practically independent of $r_g$. The redshift distortion parameter
 $\beta$, has degraded slightly, due to its strong correlation
 with $r_g$, but is still measurable at the $3 \%$ level.
 The big change from the redshifts-only analysis is in the
 galaxy-mass correlation parameter $r_g$, which is now measurable
 from the combined surveys, with a formal uncertainty of better than $2 \%$.
Comparing Figure~ \ref{pub22} with its redshift-survey-only equivalent
 Figure~\ref{pub5} it is clear that there has been a great improvement
 in the joint constraint of $\beta$ and $r_{g}$. The reason for this
 can be seen by comparing the correlations between the two parameters
 in the lower panels of both figures. In the redshifts-only analysis
 $\beta$ and $r_{g}$ are strongly anticorrelated whereas when the
 redshifts are combined with peculiar velocities this degeneracy is
 broken and the parameters become almost independent. 

\begin{figure}
\centering
\begin{picture}(200,320)
\includegraphics{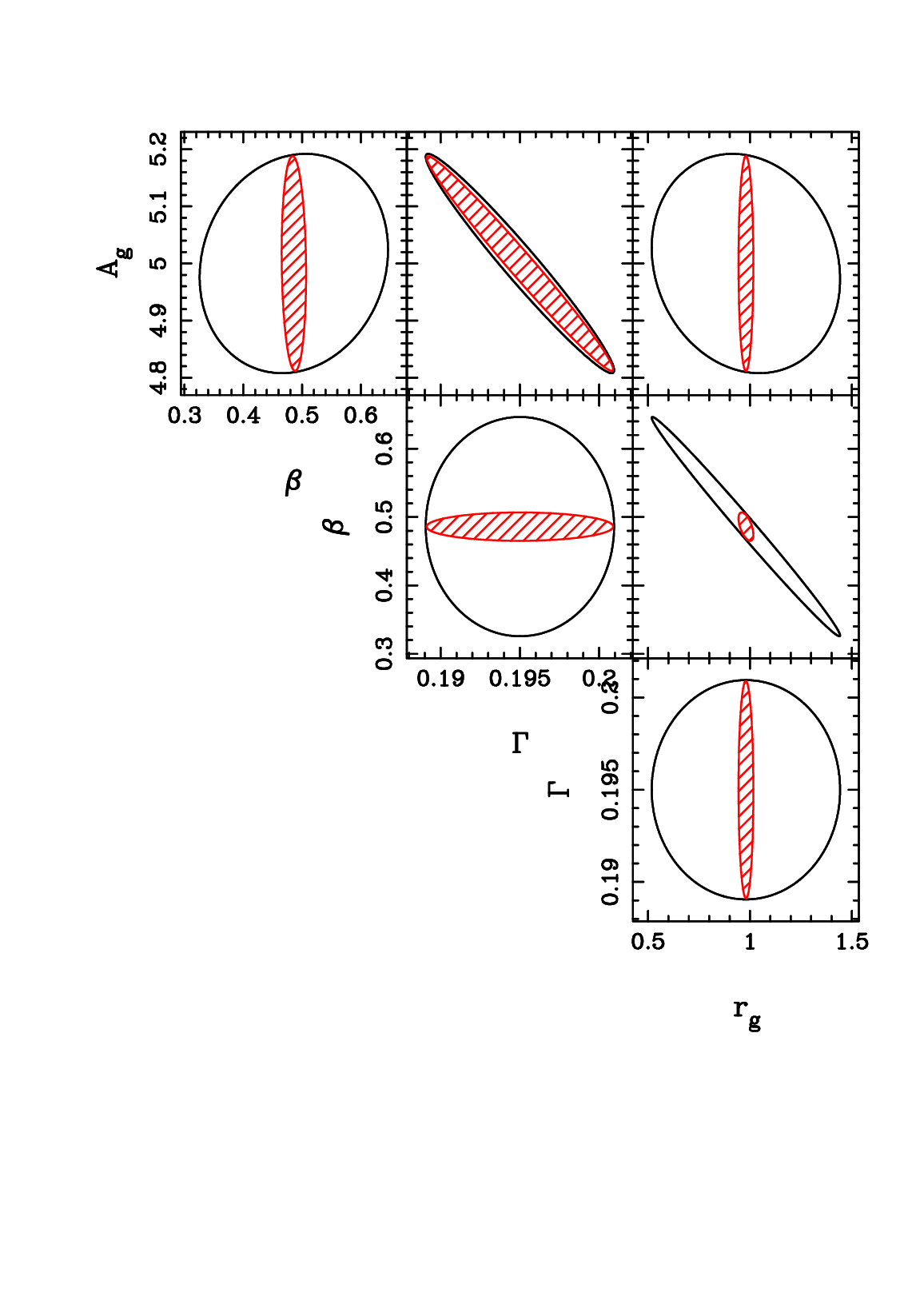}
\end{picture}
\caption{$1 \sigma$ contours for likelihood marginalised over two free
parameters.}
\label{4ellipse}
\end{figure}

Again we have plotted the expected likelihood contours for this analysis 
in order to help visualise the improvement in the constraint. In
Figure~\ref{4ellipse} the contours represent $ \Delta \ln {\rm(Likelihood)} =- \frac{1}{2}$ after marginalization over the two remaining parameters. Again the
hatched region shows the improvement from combining the data
sets. Particularly interesting is the panel
showing contours for $\beta$ and $r_{g}$ which illustrates the breaking of
the degeneracy between the two, and the subsequent improvement in their joint
constraint.

\subsection{Scale dependence of $\lowercase{b}$ and $\lowercase{r_g}$}
\begin{figure}
\centering
\begin{picture}(200,230)
\includegraphics{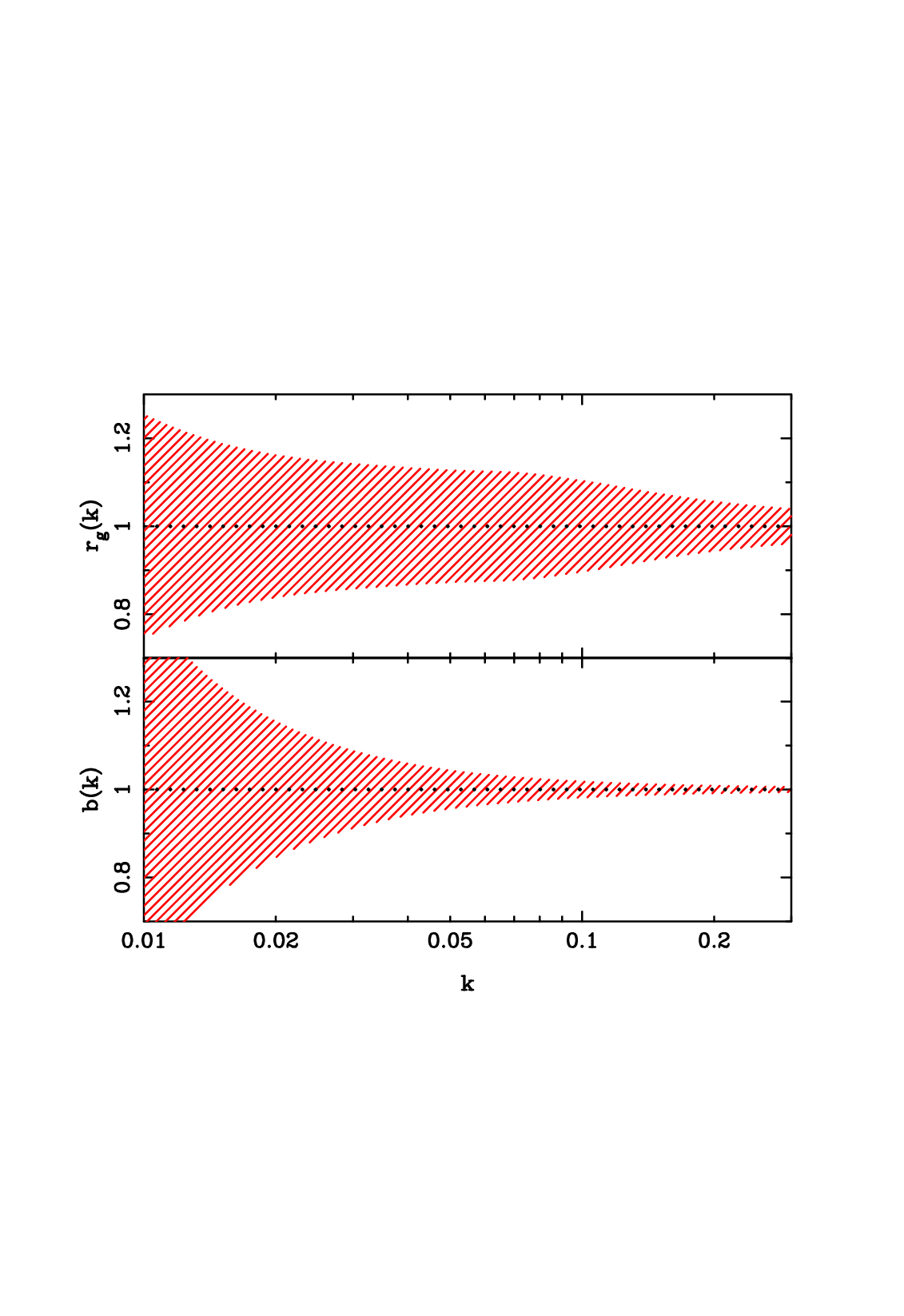}
\end{picture}
\caption{Constraining scale dependence: $\Delta b(k)$ and  $\Delta r_{g}(k)$.}
\label{scaledep}
\end{figure}

Given the tight constraint on $r_g$ and $\beta$ it is interesting to
consider a possible scale dependence in either $r_g$ or the
biassing parameter $b$. Although we have approximated $b$ to be constant
over the scale of interest, it may have some scale dependence. Another way to consider scale
dependence in the biassing is to fix $b$ as constant and allow the
parameter $r_g$ to vary with $k$. The upper panel of Figure~(\ref{scaledep})
shows the expected error bars on a measure of $r_g$ band
averaged over a logarithmic passband of width $\Delta \ln k$. The width
$\Delta \ln k$ is marked on the plot as a series of dots. $b$ is fixed at
$b=1$ for this analysis. The results are
extremely encouraging; $r_g (k)$ can be measured to within about $10\%$
over a wide range of scales improving to $4.5\%$ on the smallest
scales. The lower panel shows a similar error bar prediction for
estimates of the band -averaged $b(k)$. This plot assumes $r_g$ to be
fixed at $r_g =1$ over all scales. The bias parameter $b (k)$ is even more tightly constrained at a few percent over
a wide range.

\section{Conclusions}
 In this paper we have presented the formalism for the individual
 and combined Fisher information analysis for galaxy redshift and
 velocity field surveys. This analysis allows us to optimise both
 surveys to maximise the information content for cosmological
 parameters, providing an estimate of the uncertainty
 on the measurement of the matter and velocity power spectra and
 the set of cosmological parameters, $(A_{g},\Gamma,\beta, \omega_b,
 r_g)$. For both 6dF redshift and velocity surveys we find the optimal
 design to be as wide as possible - a result which was previously well
 known. In the case of the velocity survey we find the best design to
 be as well sampled and as accurate as possible and in the redshift
 survey we find an optimal sampling of around $70\%$. We expect to be
 able to constrain $A_g$, $\Gamma$ and $\beta$ to around $2-3\%$ from
 the redshift survey. From the velocity survey $A_v$ can be
 constrained to $5\%$ but a joint constraint of $A_v$ and $\Gamma$
 will have marginalised uncertainties of $25\%$.

We find that the major benefits of 6dF are
found when the velocity and redshift surveys are combined and when we
wish to jointly constrain the parameters $r_{g}$ and $\beta$. The
parameters' degeneracy is broken when the power
spectra are combined and the parameter $r_{g}$ can be measured much
more accurately than in any of the above surveys with just
redshifts. Finally, the scale dependence of  $r_{g}$ and $b$ can be
measured with the combined data set -- which at least will give
credance to some of the assumptions commonly made about biassing. 
Clearly the great benefit of peculiar velocity information is that it
tells us about the underlying mass and by combining this information
with galaxy redshifts we can learn much about the relationship between
luminous and dark matter.

\section*{Acknowledgements}

DB thanks the
PPARC for a studentship and ANT thanks the PPARC for an Advanced
Fellowship. We also thank Quentin Parker and Will
Saunders for invaluable help understanding the 6dF.

\section{References}
\bib
Bouchet, F.R., Prunet, S. and Sethi, S.K. 1999, MNRAS, 302, 663 
\bib
Colless, M. et al. 1999, in Looking Deep into the Southern Sky, p9
\bib
Dekel, A. and Lahav, O. 1999, ApJ, 520, 24
\bib
Eisenstein, D.J. and Hu, W. 1998, ApJ, 496, 605
\bib
Feldman, H.A., Kaiser, N. and Peacock, J. A. 1994, ApJ, 426, 23
\bib
Gregg, M. D. 1995, AJ, 110, 1052
\bib
Heavens, A. F. and Taylor, A. N. 1997, MNRAS, 290, 456l
\bib
Jorgensen, I., Franx, M. and Kjaergaard, P. 1993, ApJ, 411, 34 
\bib
Kendall, M. G. and Stuart, A. 1969, \textit{The Advanced Theory of
Statistics}, Vol 2, Griffin, London
\bib
Kochanek, C. S. et al., 2001, ApJ, 560, 566
\bib
Kolatt, T., et al. 2000, A. S. P. Conf. Ser., 201, 205 
\bib
Lawrence, C.R. and Lange A.E. 1997, AAS, 29, 1273
\bib
Peacock, J. A. and Dodds, S. J. 1994, MNRAS, 267, 1020
\bib
Peebles, P. J. E. 1980, \textit{The Large Scale Structure of the
Universe}, Princeton University Press, Princeton N.J.
\bib
Skrutskie, M. F. 2000, IAU Symposium, Vol. 204
\bib
Spergel, D. N. et al. 2003, astro-ph/0302209
\bib
Strauss, M. A. and Willick, J. A., 1995, PhR, 261, 271 
\bib
Tadros, H. et al. 1999, MNRAS, 305, 527
\bib
Taylor, A. N., Ballinger, W. E., Heavens, A. F. and Tadros, H. 2001, 
MNRAS, 327, 689
\bib
Taylor, A. N. and Watts, P. I. R. 2001, MNRAS, 328, 1027 
\bib
Tegmark, M. 1997, PhRvL, 79, 3806 
\bib
Tegmark, M., Taylor, A. N. and Heavens, A. F. 1997, ApJ 480, 22
\bib
Vogeley, M. S. and Szalay, A. S. 1996, ApJ 465, 34
\bib
Wakamatsu, K. et al. in ASP Conf. Ser., IAU 8th Asian-Pacific Regional
Metting, in press astro-ph/0306104
\bib
Zehavi, I. and Dekel, A., in ASP Conf. Ser., 201:Cosmic Flows Workshop
\bib
Zaldarriaga, M., Spergel, D.N. and Seljak, U. 1997, ApJ 488, 1

\onecolumn

\section*{Appendix A: Joint estimates of the galaxy clustering amplitude, $A_{\lowercase{g}}$,
 and the redshift-space distortion parameter, $\beta$}
\subsection*{A.1 The redshift-space Fisher Matrix}

In this Appendix we derive some exact formulae for estimating the
uncertainty on a joint measurement of the amplitude of galaxy
clustering, $A_g$, and the redshift distortion parameter, $\beta$,
in the limit of negligible shot-noise and complete correlation
between galaxies and matter, $r_g=1$, and all other parameters are
known. We model the survey volume by an effective constant volume,
$V$. The Fisher matrix in this limit is
 \be
    \eF_{ij} = \frac{1}{2} V \int \! \frac{d^3k}{(2 \pi)^3} \,
    \de_i \ln P^s_{gg}(k,\mu) \, \de_j \ln P^s_{gg} (k,\mu)
 \ee
 were $i$ and $j$ take the values $A_g$ and $\beta$.

 The galaxy redshifted power spectrum with $r_g=1$ is given by (c.f. equation
 (\ref{power}))
 \be
    P^s_{gg}(\k)
    =  D^2(k \sigma_v \mu) (1 + 2 \mu^2 \beta + \mu^4 \beta^2)
     b^2 P_{mm}(k).
 \ee
 Differentiating and angle averaging we find the Fisher matrix is
 \ba
    \eF &=&
    \left( \begin{array}{cc}
                \eF_{A_g A_g}& \eF_{\beta A_g} \\
                \eF_{\beta A_g} & \eF_{\beta \beta}
                                                        \end{array}
                                                        \right)
                                                        \nn
        &=&
    \left( \begin{array}{cc}
                2/A_g^2 &  \Phi(\beta)/A_g \beta \\
                \Phi(\beta)/A_g \beta &  \Theta(\beta)/\beta^2
                                                        \end{array}
                                                        \right)
    \frac{k^3 V}{6 \pi^2}
 \ea
 where
 \ba
    \Theta(\beta) &=&  \int^{+1}_{-1} d \mu \left( \frac{\beta \mu^2}{
            1+\beta \mu^2}\right)^2
    = 2 + \frac{1}{1+\beta}- 3 \frac{\tan^{-1} \sqrt{\beta}
    }{\sqrt{\beta}},
    \label{defA}
    \\
    \Phi(\beta) &=&  \int^{+1}_{-1} d \mu \left( \frac{\beta \mu^2}{
    1+\beta \mu^2}\right)
    =2 \left(1- \frac{\tan^{-1} \sqrt{\beta}}{\sqrt{\beta}}\right)
 \ea
 Inverting the Fisher matrix we find the parameter covariance
 matrix is
 \ba
    {\cal F}^{-1} &=&
    \left( \begin{array}{cc}
                \lgl (\Delta A_g)^2 \rgl & \lgl \Delta \beta \Delta A_g \rgl \\
                \lgl \Delta \beta \Delta A_g \rgl & \lgl (\Delta \beta)^2
                \rgl
                                                        \end{array}
                                                        \right)
                                                        \nn
        &=&
        \left( \begin{array}{cc}  A_g^2 \Theta(\beta) & - A_g \beta \Phi(\beta) \\
        - A_g \beta \Phi(\beta) & 2 \beta^2  \end{array}
                                                        \right)
        \frac{6 \pi^2}{k^3 V \Psi(\beta)}
                                                        \nn
        &=&
        \left( \begin{array}{cc} (0.01)^2 A_g^2 & - (0.02)^2 A_g \beta \\
        - (0.02)^2A_g \beta  & (0.06)^2 \beta^2
        \end{array}\right)
                \left( \frac{k}{0.2
    \Mpch}\right)^{-3} \left( \frac{V}{10^8 [\hMpc]^3}\right)^{-1}
\label{errormatrix}
 \ea
 where
 \be
    \Psi(\beta)= A_g^2 \beta^2 \, {\rm Det} \, {\cal F} = 2 \left(\frac{1}{1+\beta}
        + \frac{\tan^{-1} \sqrt{\beta}}{\sqrt{\beta}}
        - 2 \frac{[\tan^{-1} \sqrt{\beta}]^2}{\beta}\right)
 \ee
 and the approximate numerical results in the last line of
 equation (\ref{errormatrix}) are for $\beta = 0.5$.
 From this we can read off the marginalised fractional error on the amplitude
 of galaxy clustering, $\Delta A_g/A_g=0.01$, and the distortion
 parameter, $\Delta \beta/\beta =0.06$, in the absence of
 shot-noise. For a single-parameter estimate of $\beta$ the
 fractional uncertainty reduces to equation (\ref{betaerr}).

The correlation coefficient between the redshift-space distortion
parameter and the amplitude of clustering in the limit of
negligible shot-noise and all other parameters known is
 \be
    \gamma_{A_g \beta} =
    - \frac{\Phi(\beta)}{\sqrt{2 \Theta(\beta)}}
    \approx -0.78 ,
    \label{betacorr}
 \ee
 where the final numerical result is again assuming $\beta= 0.5$.

 \subsection*{A.2 The uncertainty on an estimate of the amplitude of mass clustering}
 Taylor et al (2001) and Tadros et al  (1999) proposed that the
 amplitude of mass
 clustering can be estimated from redshift surveys, if galaxies
 are perfectly correlated with mass $r_g=1$, from
 \be
    Q= (A_g \beta) \Omega_m^{-0.6}.
 \ee
 If the errors on $A_g$ and $\beta$ are similar, then this combination has the advantage that it can be estimated more
 accurately than either of its parts because it marginalises along
 the longest eigenvalue of the parameter covariance matrix. We can
 estimate the uncertainty from this combination as
 \ba
    \frac{\Delta Q}{Q} &=& \sqrt{\frac{\lgl (\Delta A_g)^2 \rgl}{A_g^2} +
                            \frac{\lgl (\Delta \beta)^2 \rgl}{\beta^2} +
                            2 \frac{\lgl \Delta A_g \Delta \beta \rgl}{A_g
                            \beta}} \nn
                        &=& \sqrt{\frac{6 \pi^2}{k^3 V}}
                        \left(\frac{\Theta(\beta) - 2 \Phi(\beta) +2}{
                        \Psi(\beta)}\right)
 \ea
 For $\beta = 0.5$ this reduces to
 \be
    \frac{\Delta Q}{Q} = 0.05 \left( \frac{k}{0.2
    \Mpch}\right)^{-3} \left( \frac{V}{10^8 (h^{-1}{\rm Mpc})^3}\right)^{-1}.
 \ee
 Hence although the projected uncertainties on $A_g$ and $\beta$
 are expected to be $1\%$ and $6\%$, the fractional uncertainty on the more
 physical parameter $Q$, can be determined to 
 $5\%$ from redshift surveys, under the assumptions
 stated above. This is a better constraint than $\beta$ but not
 $A_g$. Because the error bars in $A_g$ and $\beta$ are so different,
 the major axis of the error ellipse does not lie at $45$ degrees to
 the parameter axis and so the marginalisation does not allow Q to be
 better constrained than 
 both of the independent parameters.

\section*{Appendix B: The bivariate Fisher matrix for two correlated Gaussian fields}

 In this Appendix we explicitly calculate the Fisher matrix for
 two correlated Gaussian fields. If we label the fields X and Y the
 Fisher matrix may be written in the general form:
\begin{equation}
{\cal F}_{ij} = \int d^3 k d^3 r \sum_{{\rm XY}} \partial_{i}
C_{{\rm X}} [{\rm Cov} ( C_{{\rm X}} C_{{\rm Y}}) ]^{-1} \partial_{j}
C_{{\rm Y}}
\label{Zaldeqn}
\end{equation}
(Zaldarriaga, Spergel and Seljak 1997). In our case, X and Y can
denote $gg$, $u ^\prime u ^\prime $ and $g u ^\prime$. We can rewrite
equation~(\ref{Zaldeqn})  in a way that isolates an effective volume:
\begin{equation}
{\cal F}_{ij} = \int d^3 k  \sum_{{\rm XY}} \partial_{i}
C_{{\rm X}}  \partial_{j}
C_{{\rm Y}} \int d^3 r [{\rm Cov} ( C_{{\rm X}} C_{{\rm Y}} ) ]^{-1}
\end{equation}
The
diagonal components of the spectral covariance matrix are:
\ba
    \Cov(C_{gg}(k,r) C_{gg}(k,r)) &=& 2[P_{gg}(k)+N_{gg}(r)]^2 , \\
    \Cov( C_{u^{\prime}u^{\prime}} C_{u^{\prime}u^{\prime}} ) &=& 2[P_{u^{\prime}u^{\prime}}(k)+N_{u^{\prime}u^{\prime}}(r)]^2 ,\\
    \Cov( C_{u^{\prime}g}(k,r) C_{u^{\prime}g}(k,r) ) &=& [(P_{u^{\prime}g}(k))^{2}+(P_{u^{\prime}u^{\prime}}(k)+N_{u^{\prime}u^{\prime}}(r))
         (P_{gg}(k)+N_{gg}(r))]
\ea
with off-diagonal parts
\ba
    \Cov( C_{u^{\prime}u^{\prime}} C_{gg} ) &=&  2(C_{gu^{\prime}})^{2},\\
    \Cov( C_{u^{\prime}u^{\prime}} C_{gu^{\prime}} ) &=&  2C_{gu^{\prime}}
            (C_{u^{\prime}u^{\prime}}+N_{u^{\prime}u^{\prime}}),\\
    \Cov( C_{gg} C_{gu^{\prime}} ) &=&  2C_{gu^{\prime}}
            (C_{gg}+N_{gg}).
\ea
The full expression for the bivariate Fisher matrix is then:

\ba
   \eF_{ij} &=& \int \! \frac{d^3 k}{(2 \pi)^3}\,
     \int \!d^3\!r \,
     (C_{u^\prime u^\prime}^2  \de_i C_{gg} \de_j C_{gg} +
    2 C_{gg} C_{u^\prime u^\prime} \de_i C_{g u^\prime} \de_j C_{g u^\prime} +
    C_{gg}^2 \de_i C_{u^\prime u^\prime} \de_j C_{u^\prime u^\prime}  \nn
    & & -
     2 C_{u^\prime u^\prime} C_{g u^\prime}
     \,[\de_i C_{g u^\prime} \de_j C_{gg}  + \de_i C_{gg} \de_j C_{g u^\prime}]  \nn
     & & -
     2 C_{gg} C_{g u^\prime}
     \,[\de_i C_{g u^\prime} \de_j C_{u^\prime u^\prime}+\de_i C_{u^\prime u^\prime} \de_j C_{g u^\prime}]  \nn
     & & +
     C_{g u^\prime}^2 [\de_i C_{u^\prime u^\prime} \de_j C_{gg}  +
     2 \de_i C_{g u^\prime} \de_j C_{g u^\prime} +
     \de_i C_{gg} \de_j C_{u^\prime u^\prime}] )/
   (2 (C_{gg}  C_{u^\prime u^\prime}-C_{g u^\prime}^2)^2)
\label{supermat} \ea
where
 \ba
        C_{gg}&=& P_{gg}(k) + N_{gg}(r),\nn
        C_{u^\prime u^\prime}&=& P_{u^\prime u^\prime}(k) +
        N_{u^\prime u^\prime}(r),\nn
        C_{g u^\prime}&=& P_{u^\prime g}(k)= r_g
        \sqrt{P_{gg}(k)P_{u^\prime u^\prime}(k)}.
  \ea

\end{document}